\documentclass[12pt,useAASTEX,usenatbib,epsfig]{aastex}
\usepackage{emulateapj5}
\usepackage{epsfig}

\makeatletter

\newenvironment{apjemufigure}{%
\def\@captype{figure}%
\noindent\begin{minipage}{0.999\linewidth}\begin{center}}
{\end{center}\end{minipage}}
\makeatother

\def\l{{\ell }}

\def\healpix{H{\sc ealpix }}
\def\glesp{G{\sc lesp }}

\def\wmap{\hbox{\sl WMAP~}}

\def\etal{et al.}

\def\alm{a_{\l m}}
\def\Ylm{Y_{\l m}}
\def\Cl{C_{\l}}

\def\summ{\sum_{m=-\l}^{\l}}
\def\suml{\sum_{\l=0}^{\infty}}
\def\lm{\l m}
\def\a{{\bf a}}

\def\Ph{{\bf \Psi}}
\def\Cs{{\bf C}}
\def\Si{{\bf S}}

\def\w{{\bf w}}

\def\I{{\bf I}}

\def\S{{\bf S}}

\def\gam{{\bf \Gamma}}

\def\lp{\l^{'}}
\def\lpp{\l^{''}}

\def\planck{{\it Planck }}

\newcommand{\nbi}{{Niels Bohr Institute, Blegdamsvej 17,
DK-2100 Copenhagen, Denmark}}
\newcommand{\ru}{{Rostov State University, Space Research Department,
Zorge,5, 344091, Russia}}
\newcommand{\sao}{{Special Astrophysical Observatory, Nizhnij Arkhyz,
Karachaj-Cherkesia, 369167, Russia}}
\newcommand{\asc}{{Astro Space Center of Lebedev Physical Institute,
Profsoyuznaya 84/32, Moscow, Russia}}
\slugcomment{Submitted to {\it The Astrophysical Journal}}

\begin{document}

\title{Foreground analysis from the 1-year Wilkinson Microwave
  Anisotropy Probe (\wmap) data}
 
\author{
Pavel D. Naselsky\altaffilmark{1,2},
Oleg V. Verkhodanov\altaffilmark{3}, 
Lung-Yih Chiang\altaffilmark{1},
Igor D. Novikov\altaffilmark{1,4}
}

\altaffiltext{1}{\nbi}
\altaffiltext{2}{\ru}
\altaffiltext{3}{\sao}
\altaffiltext{4}{\asc}

\email{naselsky@nbi.dk}

\keywords{cosmology: cosmic microwave background --- cosmology:
observations --- methods: data analysis}

\begin{abstract}
We present a detailed analysis on the phases of the \wmap foregrounds
(synchrotron, free-free and dust emission) of the \wmap K-W bands in order to
estimate the significance of the variation of the spectral indices at
different components. We first extract the spectral-index varying
signals by assuming that the invariant part among different frequency
bands have 100\% cross-correlation of phases. We then use the
minimization of variance, which is normally used for extracting the
CMB signals, to extract the frequency independent signals. Such a
common signal in each foreground component could play a significant
role for any kind of component separation methods, because the methods
cannot discriminate frequency independent foregrounds and CMB.     
\end{abstract}
\section{Introduction}
The release of the first year data from the Wilkinson Microwave
Anisotropy Probe (\wmap) provides a unique opportunity to investigate
the properties of whole-sky cosmic microwave background (CMB)
anisotropies at unprecedented resolution and sensitivity to
date \citep{wmapmap,wmapsystematics,wmappw}. The \wmap science team
also produce several whole sky foreground maps of synchrotron, free-free
and dust emission from all the frequency bands at the range 23-94 GHz
with unprecedented precision \citep{wmapfg}. The analysis of the
foreground properties is important as it is an indirect testing of the
derived CMB signal properties (the angular power spectrum and its
corresponding optimal values of the cosmological parameters for the
best fit $\Lambda$CDM model \citep{toh,oliveira,efstathiou}, and it also
sheds light on the statistical properties of the
CMB signal, i.e. Gaussianity of the CMB anisotropies
\citep{wmapng,tacng,park,ndv03,ndv04,eriksenA,eriksenB,hansen}.

In the framework of inflation paradigm, the angular power spectrum $\Cl$
provides {\it all} the information about the CMB fluctuations when the
anisotropies constitute a Gaussian random field
\citep{bbks,be}. Therefore, non-Gaussianity, either with primordial
origin, or residuals from systematics or foreground component
separation, shall inflict the accuracy of the cosmological
parameters. Recently there are reports on detection of non-Gaussianity
from 1-year derived CMB maps, which is attributed to foreground residuals
 or the systematic effect correction
\citep{tacng,ndv03,ndv04,eriksenB,hansen}. This issue
therefore needs to be dealt with care before we reach any conclusions
on, among other important issues, the statistical properties of the
CMB anisotropies.

The \wmap science team use Maximum Entropy Method to produce five
maps at K-W bands for the synchrotron component, five maps for
the free-free emission. They also incorporate extrapolation of the
galactic dust emission from 100 microns \citep{dust} to \wmap
frequency range. In this paper we examine the properties of the \wmap derived
foreground maps, using an approach based on the phase of spherical
harmonic coefficients. From theoretical point of view, the
phases of the major foreground components listed above are extremely
 important. They provide an opportunity to develop and to implement so called
``blind'' method of the CMB and foregrounds separation
not for the \wmap data only, but for the upcoming high-resolution PLANCK
data as well. 

The CMB and the combined foreground signals: synchrotron, free-free, dust
emission and so on come from the same data sets. For the method by the \wmap
science team for the CMB signal extraction, that by \citet{toh} (TOH
hereafter), \citet{pcm}, and by \citet{eriksenB}, pronounced
correlations of phases (hence correlations in morphology) between different
frequencies at each foreground component are necessary
\citep{te96,bouchet,hobson,stolyarov,delabrouille,toh,patanchon,barreiro,gonzalez},
i.e. the foregrounds at different frequencies should only differ by
amplitudes not by morphology. Using the derived maps for the synchrotron, 
free-free and dust emission from the \wmap web site \footnote{{\tt
    http://lambda.gsfc.nasa.gov/product/map/m\_products.cfm}} 
we demonstrate such correlations (see Section 1). Not surprisingly,
these phases are not strongly correlated, especially for the
synchrotron emission. Such deviation in morphology is mentioned by
\wmap science team as the so-called spectral index variation effect
\citep{wmapfg}. The aim of
this paper is to find the foreground components of signal from K-W
band which cause deviations in morphology in the foregrounds (Section
2 and 3) and some components which do not have any frequency
dependence (Section 4). The latter creates some problems for the
standard tools for CMB component separation and
determines the corresponding error of the tools. Below we will
mainly discuss the synchrotron, free-free and dust components, but our
method can be easily expanded on all kind of foregrounds, which are
important for the component separation tools for the \planck mission.

This paper is arranged as follows: in Section 2 we use phase analysis
to develop a method for the spectral-index varying signals in the
foreground maps. For the \wmap foreground maps the derived
spectral-index varying signals are shown in Section 3. And in Section
4 we apply the minimization of the variance on the \wmap foreground
maps to extract the frequency independent signals for each foreground
component. We use circular statistics to test the cross correlations
of those maps in Section 5. Conclusions and discussions are in Section
6. 

\section{Separation of the spectral-index varying signal in the
  \wmap foreground maps} 
We formulate the question as follows: how can we
extract the common phases for each kind of the foregrounds from the
\wmap derived foreground maps?  

For that purpose we will apply circular phase statistics
\citep{fisher}, using Lagrangian weighting coefficients $w_j(\l)$,
($\sum_j w_j(\l)=\I$), similar to the TE96 optimization
\citep{te96,toh}, but for the phases at each band $j$. 
For statistical characterization of the temperature fluctuations on 
 a sphere we express each foreground signal as a sum over spherical harmonics
\begin{equation}
\Delta T(\theta,\varphi)=\suml \summ |\alm|e^{i\phi_{\lm}} \Ylm
(\theta,\varphi), 
\label{eq1}
\end{equation}
where $|\alm|$ and $\phi_{\lm}$ are the moduli and phases of the
coefficients of the expansion. 

In practice, we use the \healpix package \citep{healpix} to decompose
each of the \wmap foreground maps at K--W bands for the spherical
harmonic coefficients $\alm$. Then we take the {\it whitened} foreground
maps for each frequency band $j$:
\begin{equation}
M^{(j)}_{\lm}=\frac{\alm}{|\alm|}=\exp\left( i \Phi^{(j)}_{\lm} \right),
\label{map}
\end{equation}
and using the weighting coefficients $w_j(\l)$ on the phases we have the map,
for example for the synchrotron emission ${\cal S}_{\lm}$:
\begin{eqnarray} 
\lefteqn{{\cal S}_{\lm} \equiv \exp \left(i\Phi^{\cal S}_{\lm}\right)}&& \nonumber \\
&= &\Pi_{j=1}^N
  \left(M^{(j),{\cal S}}_{\lm}\right)^{w_j(\l)} = \exp \left( i
  \sum_j w_j(\l)\Phi^{(j),{\cal S}}_{\lm}\right).
\label{mapS}
\end{eqnarray}
The weighting coefficients are designed to produce the map whose phases
$\Phi^{\cal S}_{\lm}= \sum_jw_j(\l) \Phi^{(j),{\cal S}}_{\lm}$ do not depend on
frequency. The map ${\cal S}_{\lm}$ represents the common part of the
foregrounds (in morphology) among the K-W bands. Note that we can
reach different sets of weighting coefficients for the maps of the
free-free (${\cal F}_{\lm}$) and the dust (${\cal D}_{\lm}$) emissions.

\subsection{Determination of the weighting coefficients}
To obtain the weighting coefficients $w_j(\l)$ it is natural to assume that
 $\Phi_{\lm}^{\cal S}$ is orthogonal to the phase differences
 $D^{(i),{\cal S}}_{\lm}=\Phi^{(i),{\cal S}}_{\lm}-\Phi^{(5),{\cal
 S}}_{\lm}$, $i=1-5$
corresponds to the K, Ka, Q, V and W bands, respectively. Taking
 into account of $\sum_j w_j(\l)=\I$ in Eq.(\ref{mapS}) we have (for
 generality we drop the superscript ${\cal S}$)
\begin{equation}
\Phi_{\lm}=\Phi^{(5)}_{\lm}+\sum_{j=1}^4 w_j(\l) D^{(j)}_{\lm}
\label{mapS1}
\end{equation}
where $D^{(j)}_{\lm}=\Phi^{(j)}_{\lm}-\Phi^{(5)}_{\lm}$. The
orthogonality of $\Phi_{\lm}$ and $D^{(j)}_{\lm}$ leads to
\begin{equation}
\sum_{m=1}^\l \Phi_{\lm} D^{(j)}_{\lm}=0 \hspace{0.5cm} j=1, 2, 3, 4
\label{mapS4}
\end{equation}
However, considering $2\pi$ periodicity of the phases, a simple 
generalization of Eq.(\ref{mapS4}) should be considered, which can be
obtained by the following way \citep{fisher}. We introduce the
trigonometric variables
\begin{equation}
x_{\lm}=\sin\Phi_{\lm};\hspace{0.5cm} y^{(j)}_{\lm}=\sin D^{(j)}_{\lm}
\label{mapS4a}
\end{equation}
and the cross-correlation coefficients $r$ defined as follows should
produce zero value due to orthogonality, i.e.
\begin{equation}
r(\l)=\sum_{m=1}^\l \sin \Phi_{\lm} \sin D^{(k)}_{\lm} \equiv 0.
\label{mapS5}
\end{equation}
As one can see from Eq.(\ref{mapS5}), if the phase difference
$D^{(k)}_{\lm}$ and $\Phi_{\lm}$ are small $(\ll \pi/2)$, we will have
the same equation for the weighting coefficients as 
Eq.(\ref{mapS4}). If $D^{(k)}_{\lm}$ is small, but $\Phi_{\lm}$ is
  comparable with $\pi/2$ we will have non-linear equation for 
the weighting coefficients that can be solved numerically.

\subsection{Spectral-index varying foregrounds} 
We now can derive, from the given coefficients $\w$, the maps for each
component of the foregrounds (synchrotron, free-free and dust
emission) with the common phases of components among all frequency
bands as follows: 
\begin{eqnarray}
\lefteqn{\alm^{(j),{\cal S},{\cal F},{\cal D}}=\left|\alm^{(j),{\cal S},{\cal
  F},{\cal D}}\right| \exp\left(i \Phi^{{\cal S},{\cal F},{\cal
  D}}_{\lm} \right) }   \nonumber \\ 
&=&\left|\alm^{(j),{\cal S},{\cal
  F},{\cal D}}\right|  \exp \left(i\sum_j w_j(\l)
  \Phi^{(j),{\cal S},{\cal F},{\cal D}}_{\lm} \right), 
\label{comp}
\end{eqnarray}
where $j$ denotes the frequency bands, ${\cal
S}$, ${\cal F}$, ${\cal  D}$ denote the synchrotron, free-free and dust
emission, respectively, and $\Phi^{(j)}_{\lm}$ corresponds to each
component at each band.  The maps of spectral-index varying signal at
each band for each foreground component are  
\begin{eqnarray}
\lefteqn{\alm^{(j),{\cal S},{\cal F},{\cal D}}=\left| \alm^{(j),{\cal
        S},{\cal F},{\cal D}}  \right| }&& \nonumber \\
&& \left[\exp\left(i
  \Phi^{(j),{\cal S},{\cal F},{\cal D}}_{\lm}\right) - \exp \left( i \sum_j
  w_j(\l)\Phi^{(j),{\cal S},{\cal F},{\cal D}}_{\lm}\right) \right].
\label{comp}
\end{eqnarray}
Taking into account Eq.(\ref{comp}) one gets the
map of fluctuation 
\begin{equation}
\alm^{(j),{\cal S},{\cal F},{\cal D}} \simeq \left| \alm^{(j),{\cal
      S},{\cal F},{\cal D}}\right| 
\Delta^{(j)}_{\lm}\exp\left[i\left(\Phi^{(j),{\cal S},{\cal F},{\cal
    D}}_{\lm}+\frac{\pi}{2}\right)\right],  
\label{fluc}
\end{equation}
where $\Delta^{(j)}_{\lm}=\Phi^{(j)}_{\lm}-\Phi_{\lm}$. The map from Eq.(\ref{fluc}) represents all variation of the phase for
all $m \ge 1$ harmonics of the signal, if $\Delta^{(j)}_{\lm} \ll
1$. As one can see, for that map the moduli are
$|\alm^{(j),{\cal S},{\cal F},{\cal D}}||\Delta^{(j)}_{\lm}|$ and the
phases are the initial phases rotated by the angle $\pi/2$, if
$\Delta^{(j)}_{\lm}>0$  and by the angle
$3\pi/2$, if $\Delta^{(j)}_{\lm} < 0$.

\section{The spectral-index varying signals in the \wmap
  synchrotron, free-free and dust maps} 
We implement the method we have developed to extract
the peculiar signals for the synchrotron,
free-free and dust emission for the \wmap frequency bands. We do not
include the K band in our analysis due to the same reason as the \wmap
science team mentions. The maps for the spectral-index varying signals
are grouped in three: synchrotron, free-free and dust emission. 

In Fig.\ref{syn} we plot the maps for the spectral-index varying signal of the
synchrotron emission for the \wmap Ka-W bands. The $m=0$
modes in these spectral-index varying maps have zero amplitude. It is
the reason why synchrotron maps looks peculiar.

Fig.\ref{free} and \ref{dust} are the spectral-index varying signals
for the free-free  and dust emission, respectively. Again the $m=0$
modes are naturallly excluded.

\begin{apjemufigure}
\hbox{\hspace*{-0.2cm}
\centerline{\includegraphics[width=1.\linewidth]{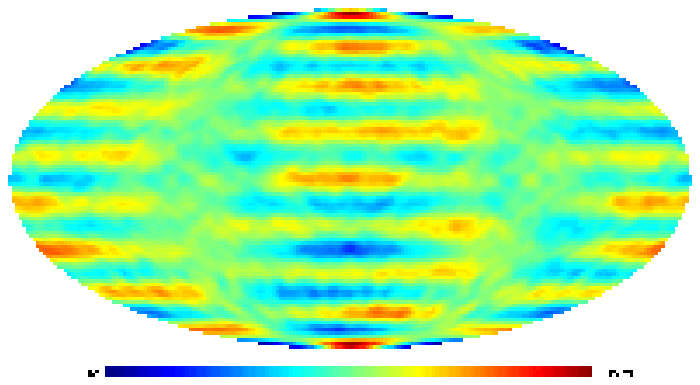}}}
\hbox{\hspace*{-0.2cm}
\centerline{\includegraphics[width=1.\linewidth]{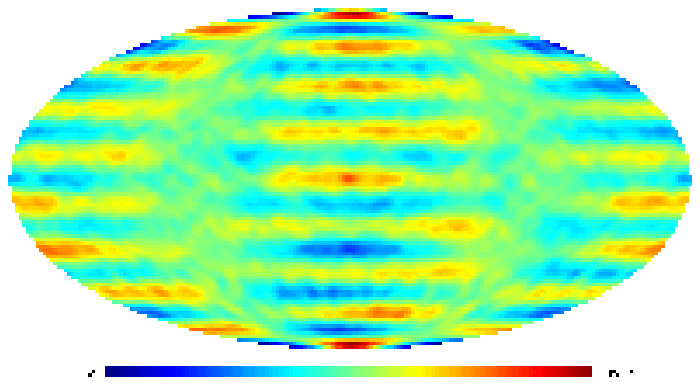}}}
\hbox{\hspace*{-0.2cm}
\centerline{\includegraphics[width=1.\linewidth]{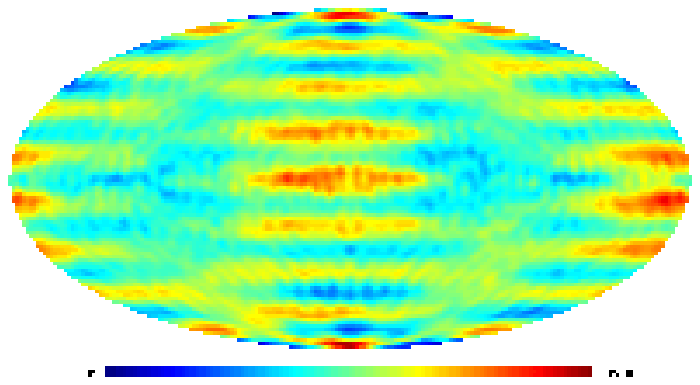}}}
\hbox{\hspace*{-0.2cm}
\centerline{\includegraphics[width=1.\linewidth]{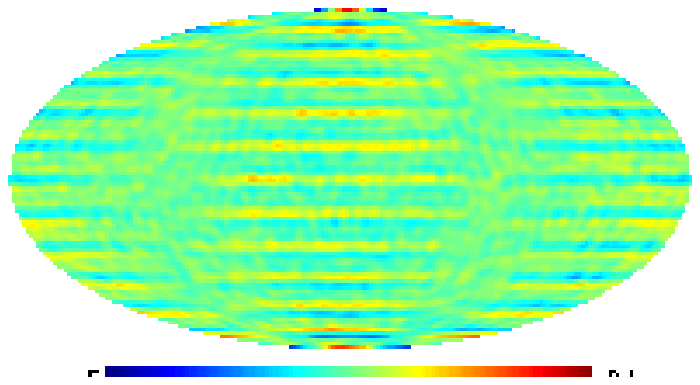}}}
\caption{The derived spectral-index varying signals for
the synchrotron component. From top to bottom are Ka, Q, V and W
bands.}
\label{syn} 
\end{apjemufigure}

For the \wmap dust maps the spectral-index varying signals are relatively
weak in comparison with those from the synchrotron and free-free maps. 
However, it is in order of 30\% of the amplitude of ILC signal which
should be taken into account seriously. 
 
In Fig.\ref{sfd} we display the spectral-index varying
signals, the sum of the synchrotron, free-free and dust
emission.

\begin{apjemufigure}
\hbox{\hspace*{-0.2cm}
\centerline{\includegraphics[width=1.\linewidth]{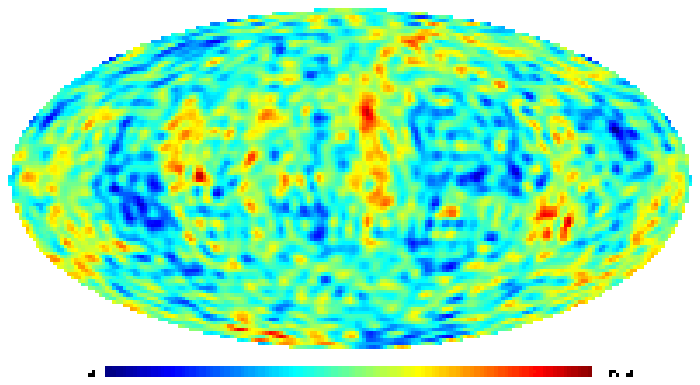}}}
\hbox{\hspace*{-0.2cm}
\centerline{\includegraphics[width=1.\linewidth]{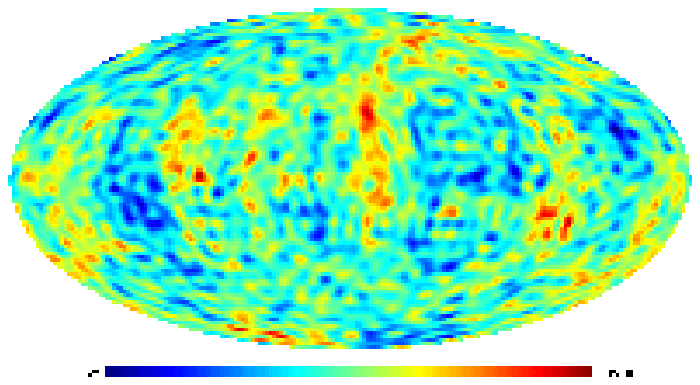}}}
\hbox{\hspace*{-0.2cm}
\centerline{\includegraphics[width=1.\linewidth]{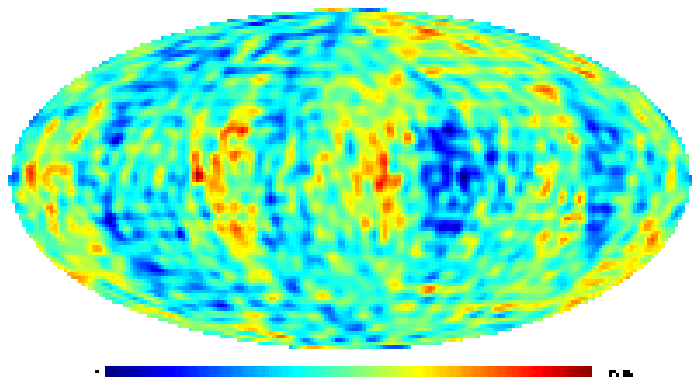}}}
\hbox{\hspace*{-0.2cm}
\centerline{\includegraphics[width=1.\linewidth]{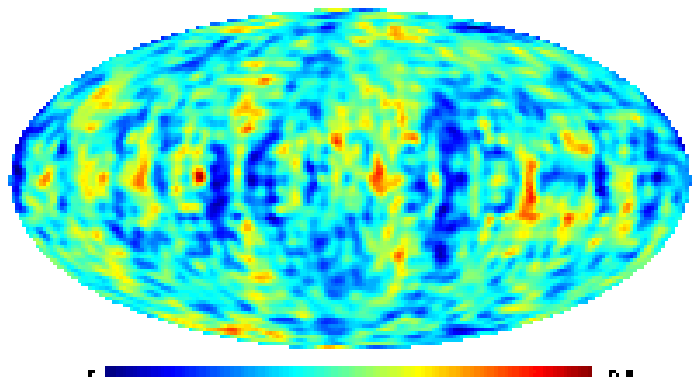}}}
\caption{The derived spectral-index varying signals for the
free-free emission. From top to bottom are Ka, Q, V and W bands.}
\label{free} 
\end{apjemufigure}

\begin{apjemufigure}
\hbox{\hspace*{-0.2cm}
\centerline{\includegraphics[width=1.\linewidth]{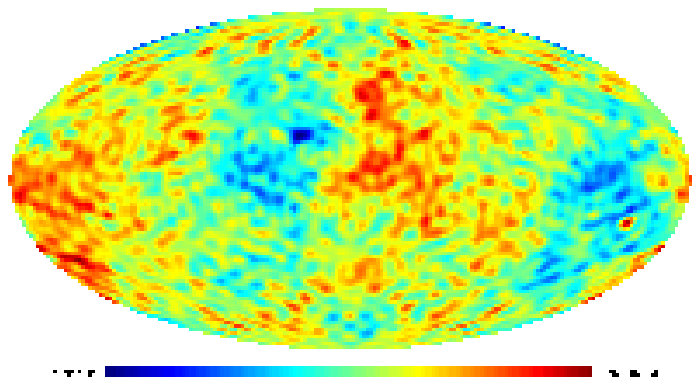}}}
\hbox{\hspace*{-0.2cm}
\centerline{\includegraphics[width=1.\linewidth]{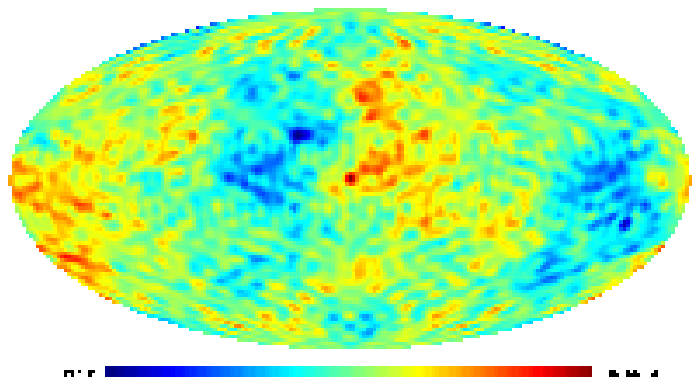}}}
\hbox{\hspace*{-0.2cm}
\centerline{\includegraphics[width=1.\linewidth]{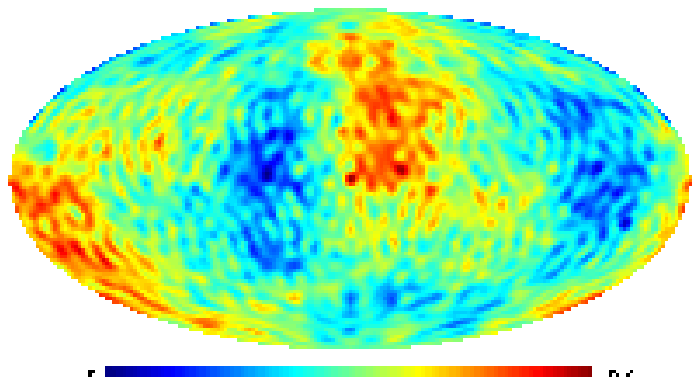}}}
\hbox{\hspace*{-0.2cm}
\centerline{\includegraphics[width=1.\linewidth]{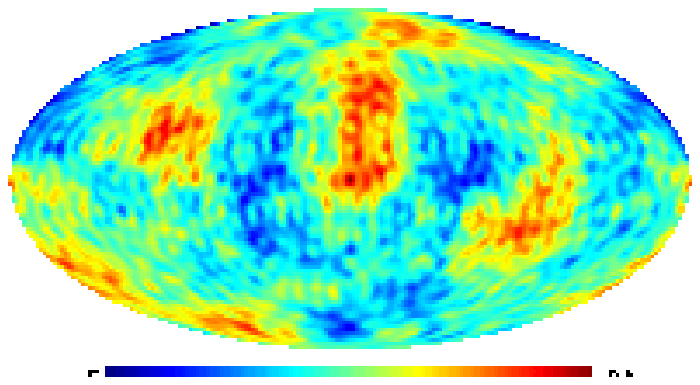}}}
\caption{The derived spectral-index varying signals for the dust
emission. From top to bottom are Ka, Q, V and W bands.}
\label{dust} 
\end{apjemufigure}

\begin{apjemufigure}
\hbox{\hspace*{-0.2cm}
\centerline{\includegraphics[width=1.\linewidth]{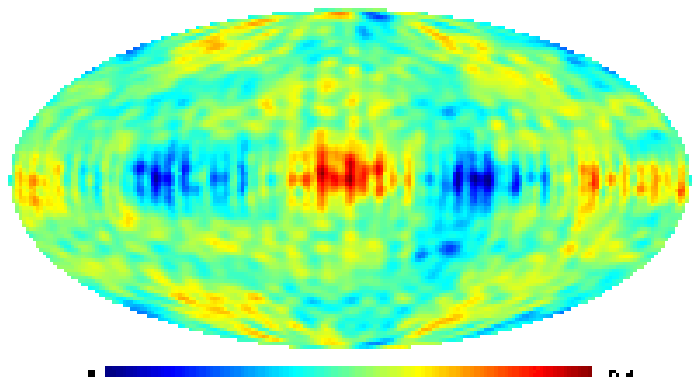}}}
\hbox{\hspace*{-0.2cm}
\centerline{\includegraphics[width=1.\linewidth]{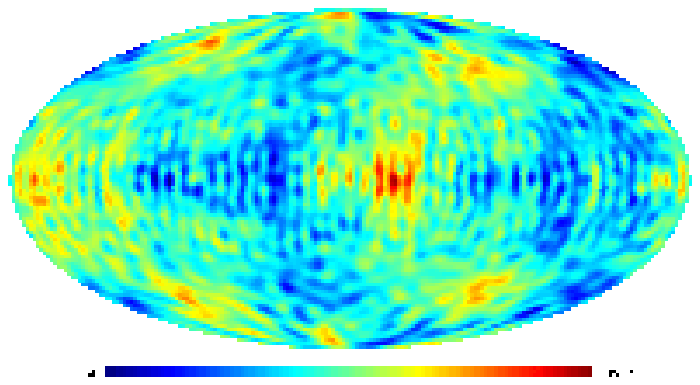}}}
\hbox{\hspace*{-0.2cm}
\centerline{\includegraphics[width=1.\linewidth]{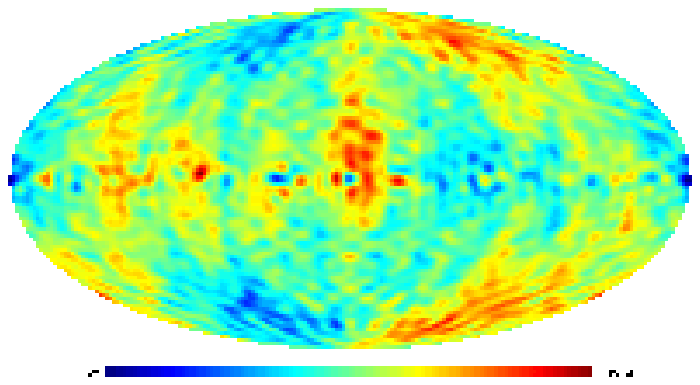}}}
\hbox{\hspace*{-0.2cm}
\centerline{\includegraphics[width=1.\linewidth]{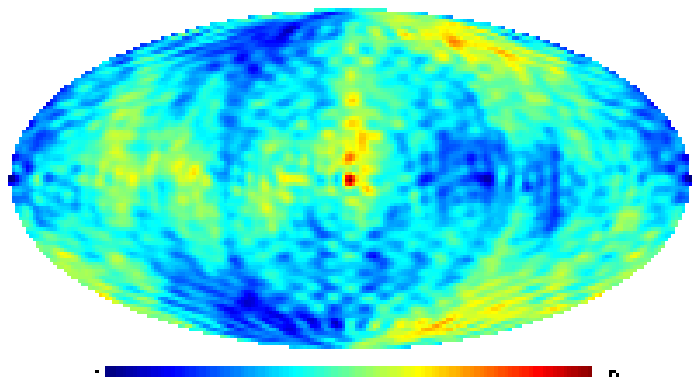}}}
\caption{The derived spectral-index varying signals of the sum
  of the synchrotron, free-free and dust emissions. From top to bottom
  are ka, Q, V and W bands.} 
\label{sfd} 
\end{apjemufigure}

\section{The frequency independent signals (FIS) in the \wmap foreground maps}
In this Section we use the minimizatoin of the variace in a similar
fashion on the \wmap foreground maps for the frequency independent
signals (hereafter FIS). This method
is mainly used by the \wmap science team to produce the CMB map
\citep{wmapfg} using a weighted combination of the 5 bands outside the
Galactic plane. \citet{toh} (hereafter TOH) perform an independent
foreground analysis from the \wmap data using weighting
coefficients dependent not only on the angular scales ($\ell$) but also on
the Galactic latitudes. Below we use the TOH method for
the synchrotron, free-free and dust maps in order to extract
the frequency independent component (or partly independent for some of the
frequency range) without any assumptions about the statistical nature of
the derived signals. Moreover, we use the whole sky for analyses and
do not apply any Galaxy plane cut-off and masks, nor do we dissect the
whole sky into disjoint regions. 

Using minimization scheme by TOH and \citet{te96}, we implement the
blind method for the FIS for synchrotron, free-free
and dust emission which have the same
phases for K-W bands. Unlike Section 1, here we are looking for some
common component of the foregrounds, which is frequency independent
at the \wmap frequency range. Moreover, such a common component could
play a significant role for any kind of component separation methods,
because the methods cannot discriminate frequency independent
foregrounds and CMB.    
 
\subsection{The synchrotron FIS maps} 
We firstly group the maps in the following combinations for the \wmap
synchrotron maps: Ka--Q, Ka--V and Q--V. 
For each pair of the maps (Ka--Q, Ka--V and Q--V) we
want to reach the common Synchrotron emission map ($\alm^S \equiv \S$) which
has the same phases for each pair of the bands and for each
$\l,m$ mode. We use the
optimization coefficients ${\gam}^{(i)} \equiv \Gamma_{\ell}^{(i)}$,
similar to that by TE96 \citet{te96}, TOH, \citet{pcm}:
\begin{equation}
\S=\alm^S=\sum_{i=1}^2
\Gamma_{\ell}^{(i)}\alm^{(i)} \equiv \sum_{i=1}^2{\gam}^{(i)}\a^{(i)},
\label{eq14}
\end{equation}
where $i=1,2$ correspond to the two bands in a pair.
The $\gam^{(i)}$ coefficients are subject to the following constraints
\begin{equation}
\frac{\delta \summ|\S|^2} {\delta \gam^{(i)}}  =  0,
\label{minvariance} 
\end{equation}
and
\begin{equation}
  \sum_i\gam^{(i)}  =  \I, 
\label{eq15}
\end{equation}
where $\delta/\delta \gam^{(i)}$ denotes functional derivatives.

The $\gam^{(i)}$ then have the following forms
\begin{eqnarray}
\gam^{(1)}& =\frac{\summ \left\{|\a^{(2)}|\left[\,|\a^{(2)}|-
    |\a^{(1)}|\cos(\Ph^{(2)}-\Ph^{(1)})\,\right]\right\}}
    {\summ|\a^{(1)}-\a^{(2)}|^2},  \nonumber \\
\nonumber \\
\gam^{(2)} &=\frac{\summ\left\{|\a^{(1)}|\left[\,|\a^{(1)}|-|\a^{(2)}|
\cos(\Ph^{(2)}-\Ph^{(1)})\,\right]\right\}}
{\summ|\a^{(1)}-\a^{(2)}|^2}, 
\label{eq16}
\end{eqnarray}
where $\Ph^{(i)}$ are the phases of the synchrotron signal $\a^{(i)}$ at
the $i$-th map.

Then we obtain the common signal $\S$ and the frequency dependent part
$F^{(j)}_{\lm}$ at each $j$-th bands.  
\begin{eqnarray}
F^{(1)}_{\lm}=\a^{(1)}-\S= \gam^{(2)}(\a^{(1)}-\a^{(2)}) \nonumber \\
F^{(2)}_{\lm}=\a^{(2)}-\S= -\gam^{(1)}(\a^{(1)}-\a^{(2)}).
\label{eq16a}
\end{eqnarray}

\begin{apjemufigure}
\hbox{\hspace*{-0.2cm}
\centerline{\includegraphics[width=1.\linewidth]{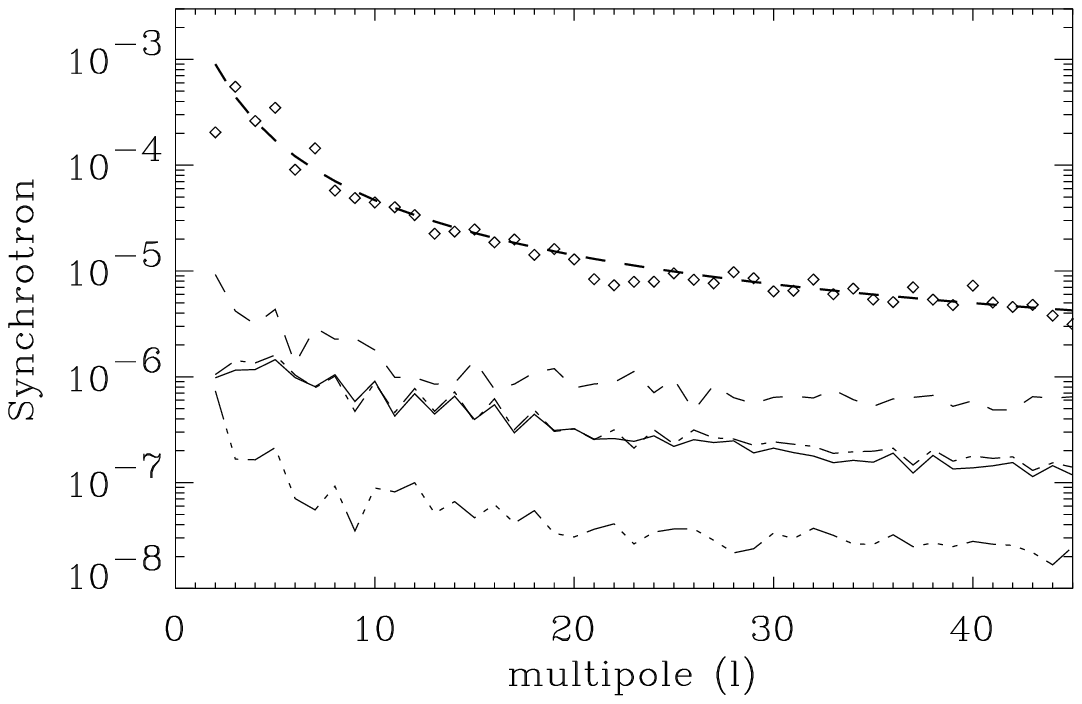}}}
\caption{The power spectra of the common component
from the synchrotron emission. Top thick dash line corresponds to
the \wmap best fit $\Lambda$CDM model. Boxes reproduce the ILC power
spectrum. Thin long dash line is the common component from Ka--Q
synchrotron maps. Solid line is the common component from Ka--V
synchrotron maps. Dash-dot line is
that from Q--V and Dash dot-dot line is that from V--W.  
}
\label{Spow} 
\end{apjemufigure}
In Fig.\ref{Spow} we plot the power spectra for the FIS taken from
different combinations of the \wmap synchcrotron maps. As one can see,
for Q--V bands and Ka--V bands 
we have the frequency independent component. 
 
\begin{apjemufigure}
\hbox{\hspace*{-0.2cm}
\centerline{\includegraphics[width=1.\linewidth]{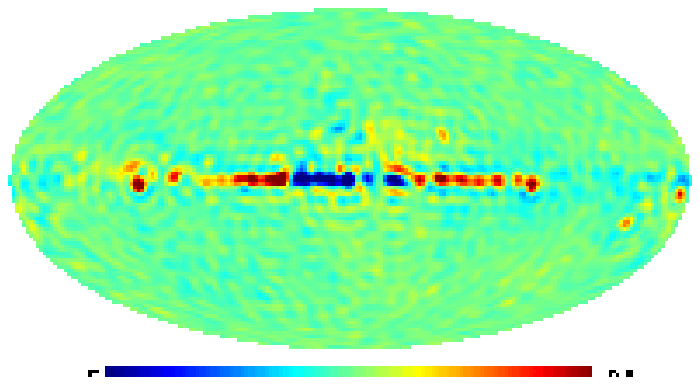}}}
\caption{The FIS from Q--V synchrotron maps.  
}
\label{Smap} 
\end{apjemufigure}

In Fig.\ref{Smap} we show the FIS from
Q--V synchrotron maps, which mainly concentrated in the Galactic
plane, but some of the point source residues seen above and below
it. The power spectra of such FIS shown in Fig.\ref{Spow} is
smaller than the ILC power, as shown in Fig.\ref{Spow}. However, due
to the  source clustering in the Galactic plane it can significantly
perturb the CMB map at Galactic plane region (see for example,
\citet{pcm}).

\begin{apjemufigure}
\hbox{\hspace*{-0.2cm}
\centerline{\includegraphics[width=1.\linewidth]{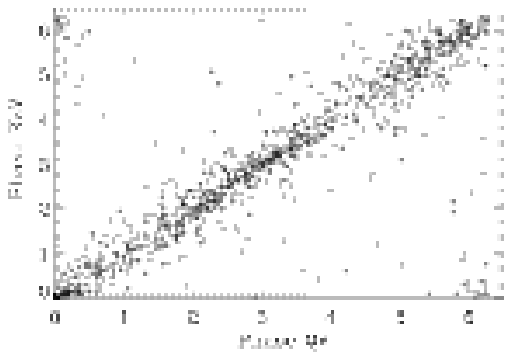}}}
\hbox{\hspace*{-0.2cm}
\centerline{\includegraphics[width=1.\linewidth]{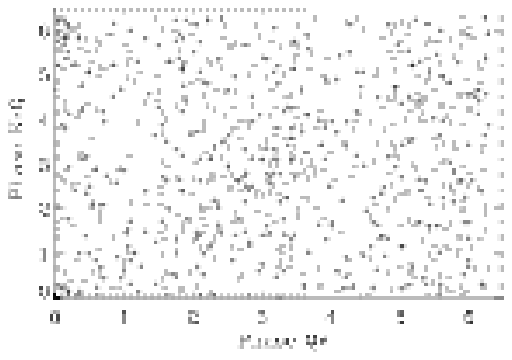}}}
\hbox{\hspace*{-0.2cm}
\centerline{\includegraphics[width=1.\linewidth]{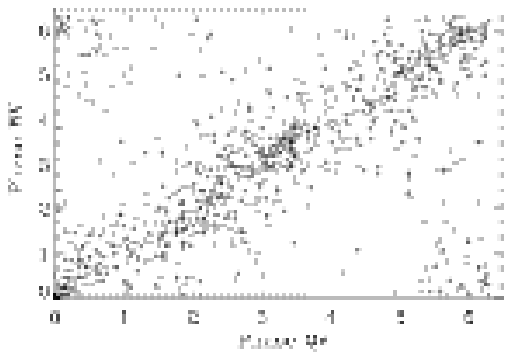}}}
\hbox{\hspace*{-0.2cm}
\centerline{\includegraphics[width=1.\linewidth]{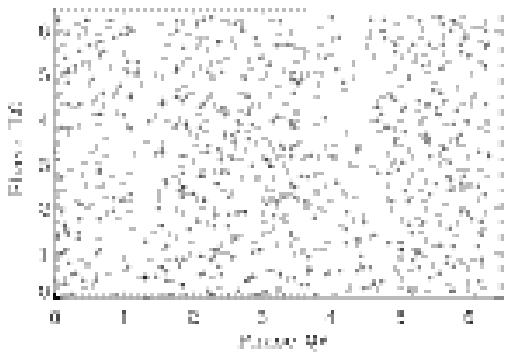}}}
\caption{Reconstructed phases of the FIS from synchrotron maps with
  different combination of the bands. From top to bottom panels are
  cross-correlation between combination of Ka--V ($y$-axis) against
  Q--V ($x$-axis), Ka--Q against Q--V, V--W against Q--V, and ILC against Q--V. }
\label{phasS} 
\end{apjemufigure}

In Fig.\ref{phasS} we plot the cross-correlations of phases
for the synchrotron FIS obtained at different frequency
bands. This Figure confirms that the FIS is common for
Ka--V bands.

\subsection{The dust FIS maps}
We show the dust FIS, using the same separation technique as for
synchrotron emission. 

\begin{apjemufigure}
\hbox{\hspace*{-0.2cm}
\centerline{\includegraphics[width=1.\linewidth]{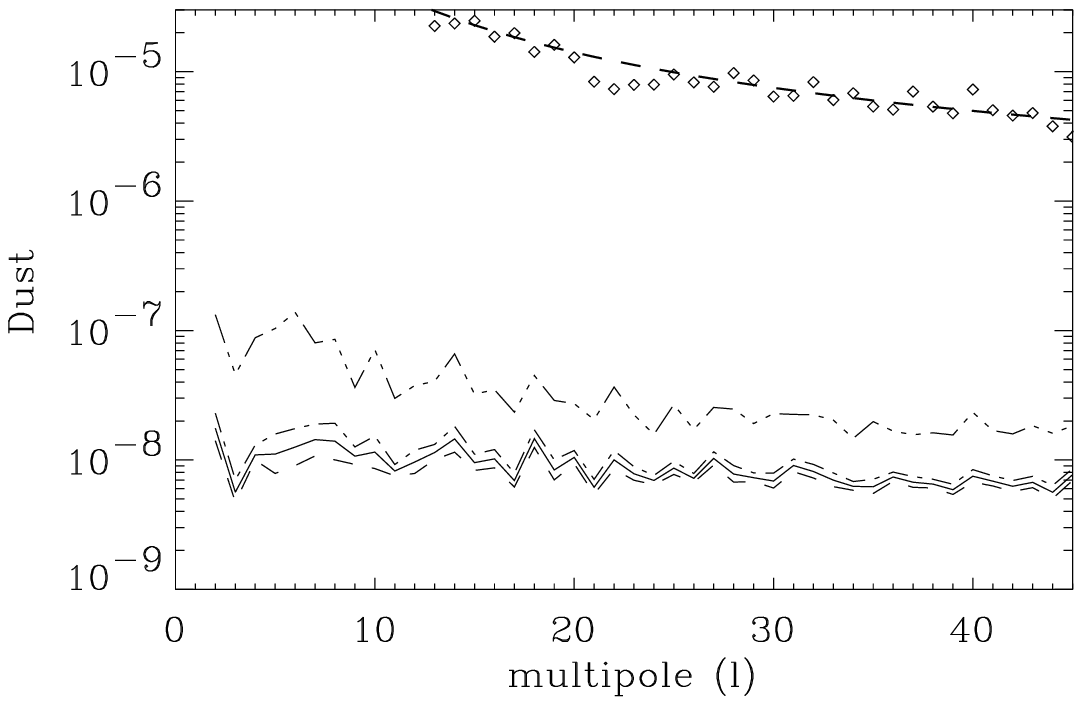}}}
\caption{The power spectra of the common component
from the dust emission. Top thick dash line corresponds to
the \wmap best fit $\Lambda$CDM model. Boxes reproduce the ILC power
spectrum. Thin long dash line is the common component from Ka--Q
synchrotron maps. Solid line is the common component from Ka--V
synchrotron maps. Dash-dot line is
that from Q--V and Dash dot-dot line is that from V--W.
}
\label{Dpow} 
\end{apjemufigure}

\begin{apjemufigure}
\hbox{\hspace*{-0.2cm}
\centerline{\includegraphics[width=1.\linewidth]{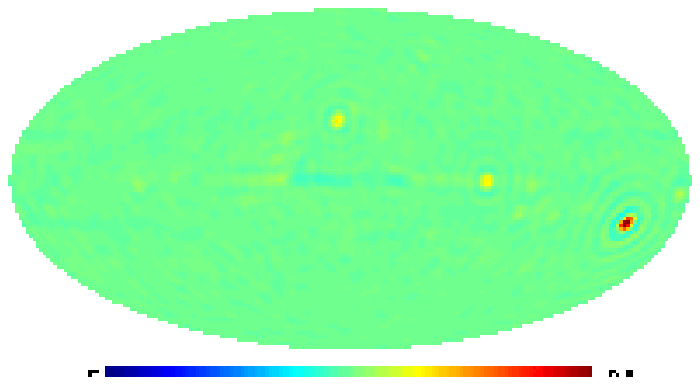}}}
\caption{The FIS from Q-V dust maps.  
}
\label{Dmap} 
\end{apjemufigure}

Fig.\ref{Dpow} is the power spectra. As shown in Fig.\ref{Dmap} the
 FIS is mainly related to the point sources. Contamination of the
 Galactic plane signal is week. In Fig.\ref{phdust} we plot the
cross correlation of phases for the FIS at different combinations of
the bands. These correlations are very strong for all Ka--V bands.

\begin{apjemufigure}
\hbox{\hspace*{-0.2cm}
\centerline{\includegraphics[width=1.\linewidth]{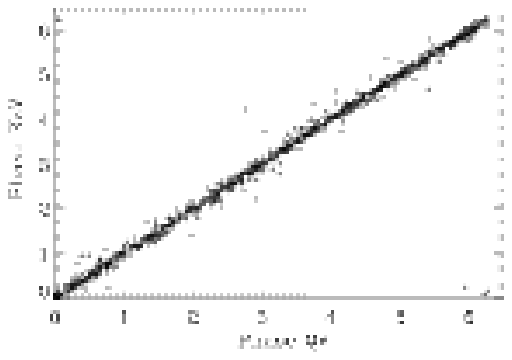}}}
\hbox{\hspace*{-0.2cm}
\centerline{\includegraphics[width=1.\linewidth]{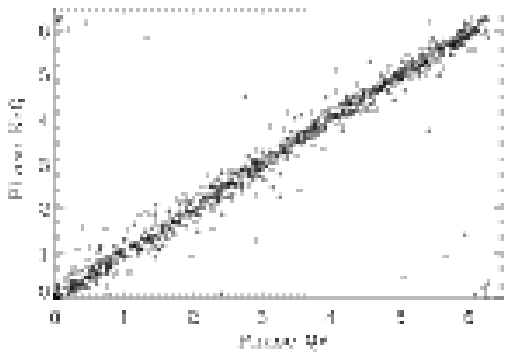}}}
\hbox{\hspace*{-0.2cm}
\centerline{\includegraphics[width=1.\linewidth]{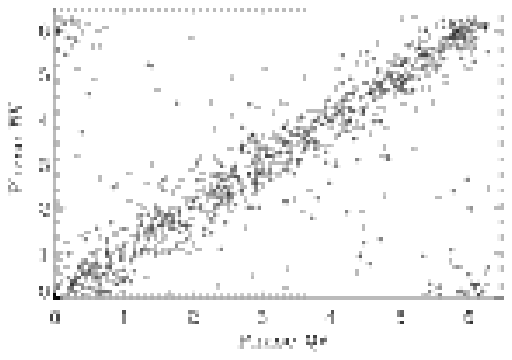}}}
\hbox{\hspace*{-0.2cm}
\centerline{\includegraphics[width=1.\linewidth]{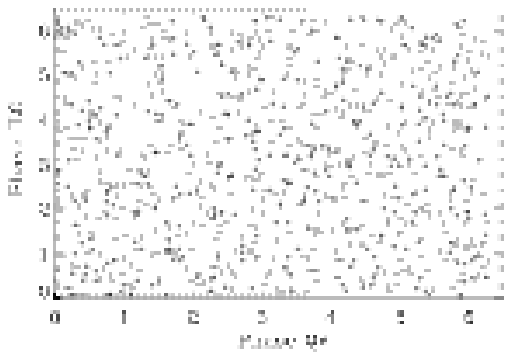}}}
\caption{Reconstructed phases of the FIS from the dust maps with
  different combination of the bands. From top to bottom panels are
  cross-correlation between combination of Ka--V ($y$-axis) against
  Q--V ($x$-axis), Ka--Q against Q--V, V--W against Q--V, and ILC
  against Q--V.}
\label{phdust} 
\end{apjemufigure}

\begin{apjemufigure}
\hbox{\hspace*{-0.2cm}
\centerline{\includegraphics[width=1.\linewidth]{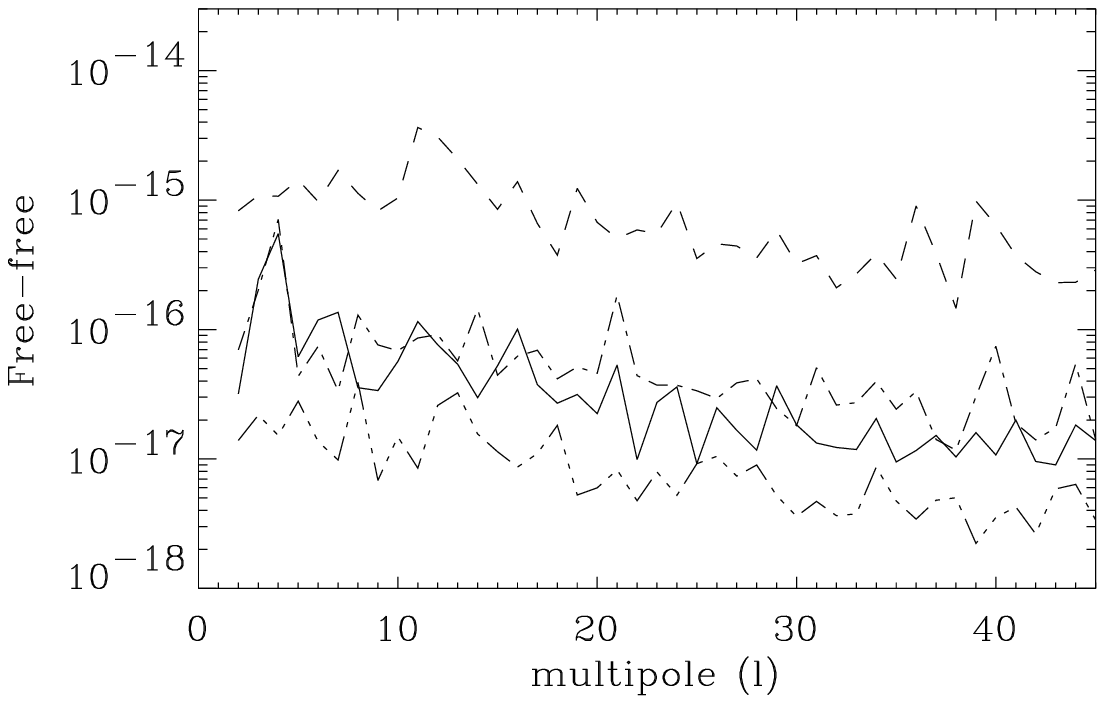}}}
\caption{Power spectra for the common component from free-free
  emission. The definition of the lines are the same as in Fig.\ref{Spow}}
\label{ffpow} 
\end{apjemufigure}
The power spectra are different for different frequency bands and phases of the
signal are not highly correlated. Actually, the free-free component of
the \wmap foreground is the only one which does not have FIS.

\subsection{The free-free FIS maps}
For the free-free emission the derived FIS is extremely weak, as
one can see from the Fig.\ref{ffpow}.

\section{Cross Correlation of phases between the derived CMB maps and the
  foregrounds}
The aim of this Section is to show how the variation of the spectral
index in the foreground components can manifest themselves in the CMB
signal, which should be clean from the foreground contamination. We
use the whole-sky maps derived from the \wmap data using different
separation methods. Namely we use the \wmap Internal Linear
Combination (ILC) map, the foreground cleaned map (FCM) and Wiener
filtered map (WFM) by TOH, and the phase cleaned map (PCM) by
\citet{pcm}.

We use the trigonometric moment statistics to counter the circular nature of
phases (\citet{fisher}, see also \citet{ndv03}). We define the
following trigonometric moments:
\begin{eqnarray}
\Cs(\l,\Delta \l) &=& \frac{1}{\l-\lp} \sum_{m=1}^{\l-\lp}
        \cos(\Phi_{\l-\lp,m}-\Psi_{\l+\lpp,m}); \nonumber \\
\Si(\l, \Delta \l) &=& \frac{1}{\l-\lp} \sum_{m=1}^{\l-\lp}
        \sin(\Phi_{\l-\lp,m}-\Psi_{\l+\lpp,m}); \nonumber \\
r^2(\l,\Delta \l) &=& \left[ \Cs^2(\l,\Delta \l) +\Si^2(\l,\Delta \l)
        \right] \l ,
\label{def2}
\end{eqnarray}
where $\Delta \l = \lp + \lpp$ and $\Psi_{\lm}$ are the phases
of the foreground-cleaned signals (ILC, FCM, WFM, PCM) at each $\l m$
harmonic and $\Phi_{\lm}$ are the phases of the foregrounds. As we
have shown, the most dangerous component of the foreground at each
frequency band is the spectral index varying tail in the synchrotron,
free-free and dust emission, shown in Fig. Below we plot the cross
correlation function $\Si$, $\Cs$ and $r^2$ for each of the foreground
cleaned signals. Using the statistics we look for signatures that
deviate from the 68\% CL. If the phases of the cleaned map are highly
correlated with those of the foregrounds, the $\Cs$ statistic has
asymptotics to unity, whereas the $\Si$ to zero. For the case of
non-correlations, both $\Cs$ and $\Si$ statistics have $\l^{1/2}$
asymptotics due to ``random walk''.

\begin{figure*}

\epsfig{file=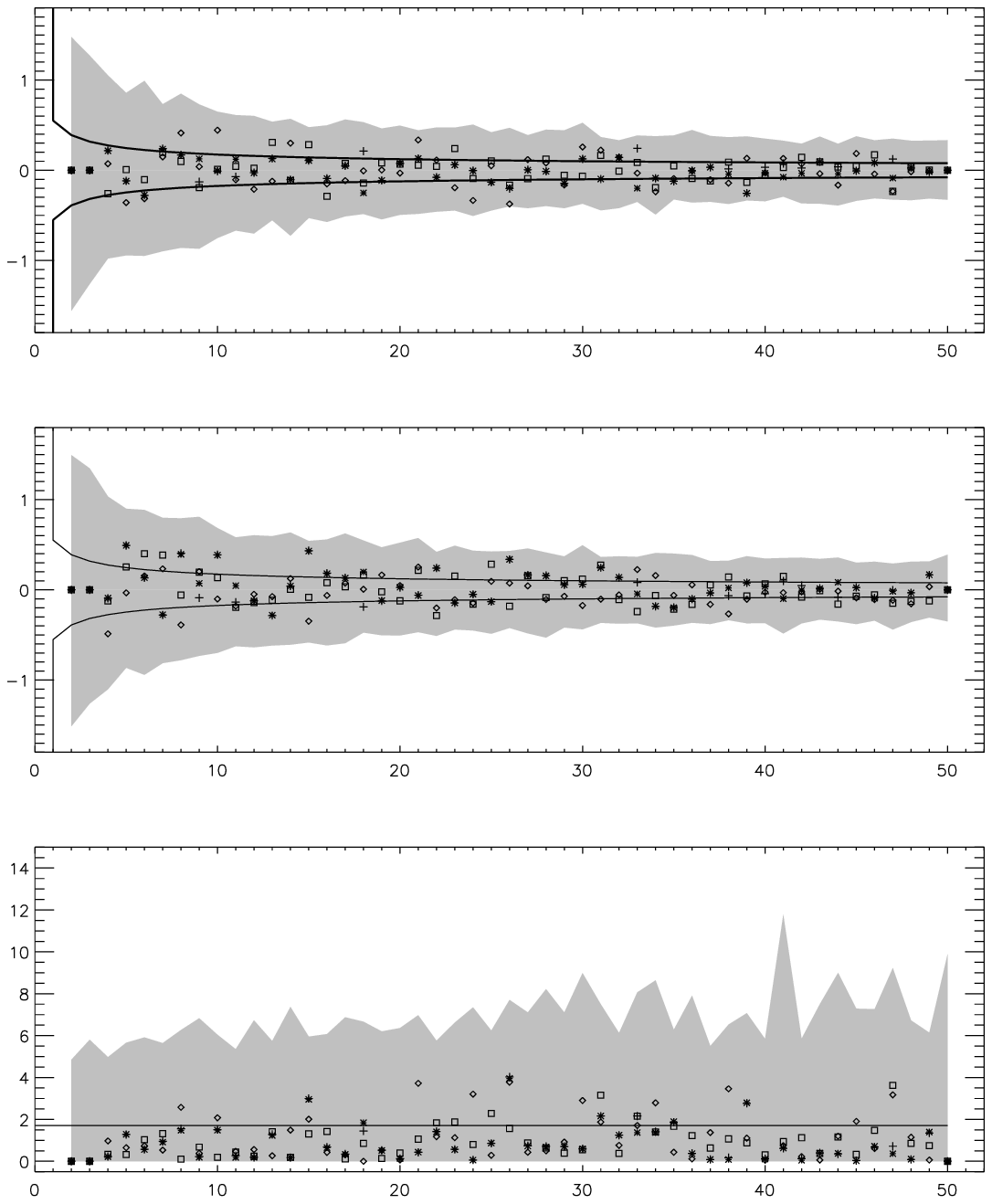,width=6.2cm,height=11cm}
\epsfig{file=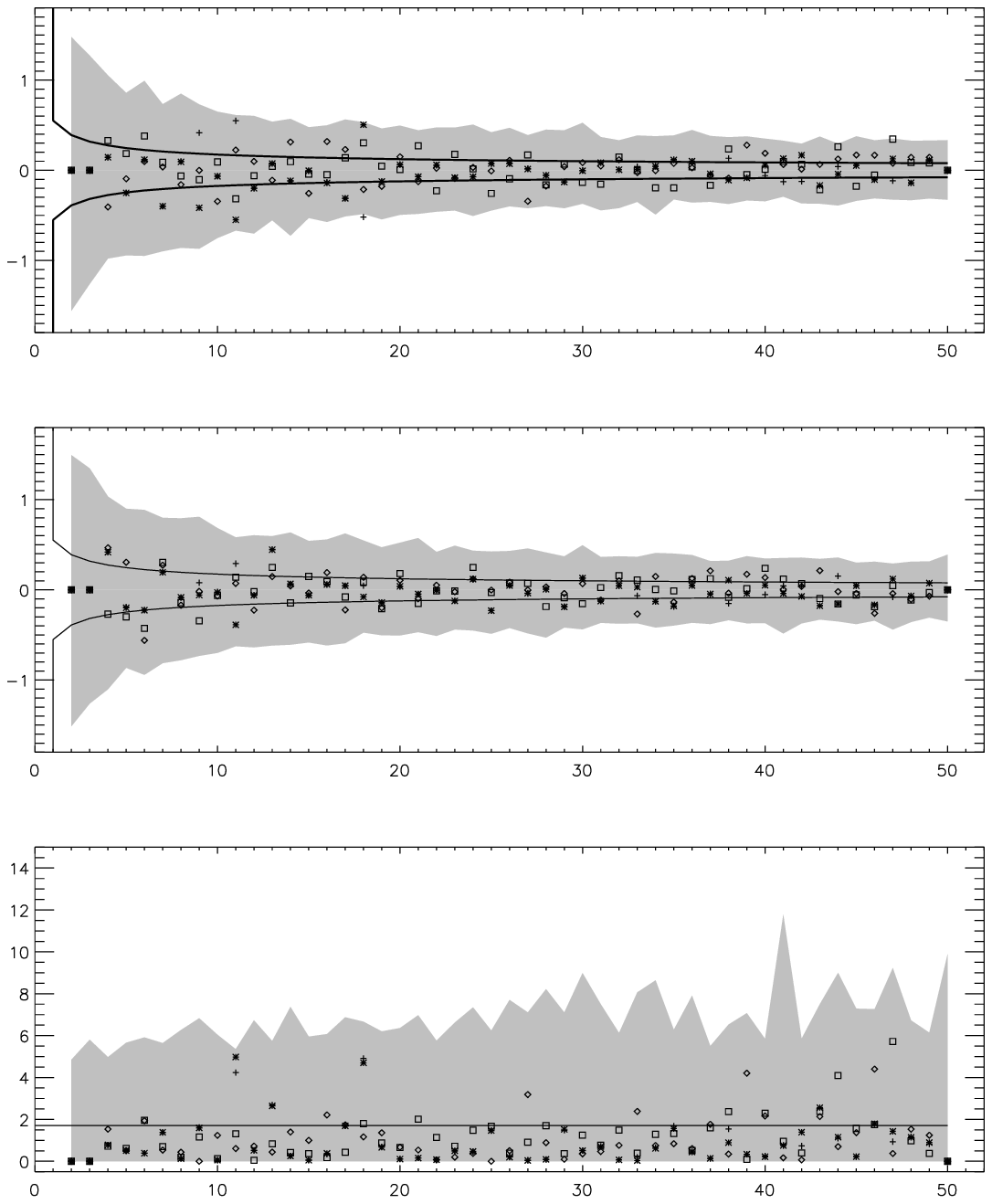,width=6.2cm,height=11cm}
\epsfig{file=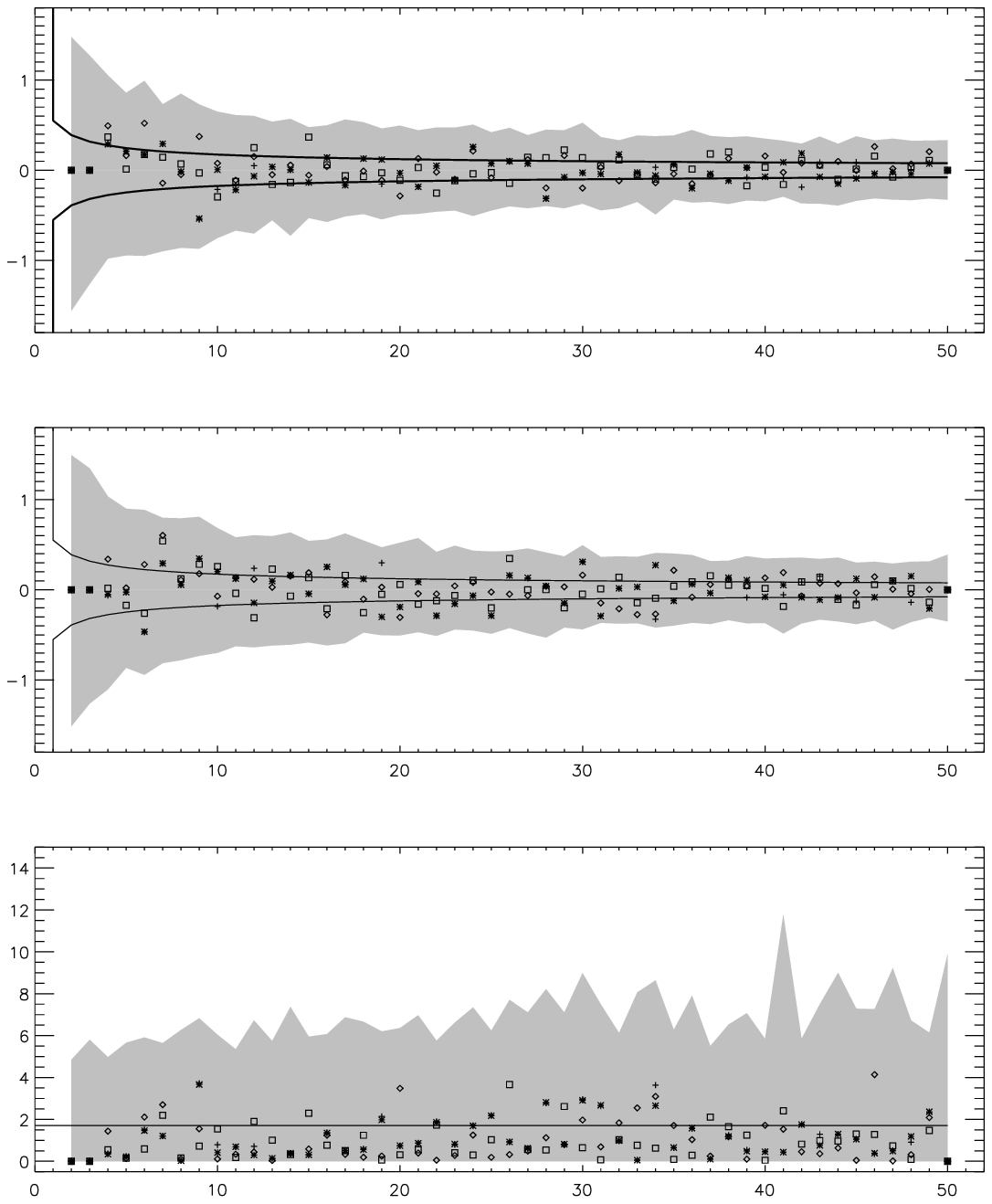,width=6.2cm,height=11cm}\\
\epsfig{file=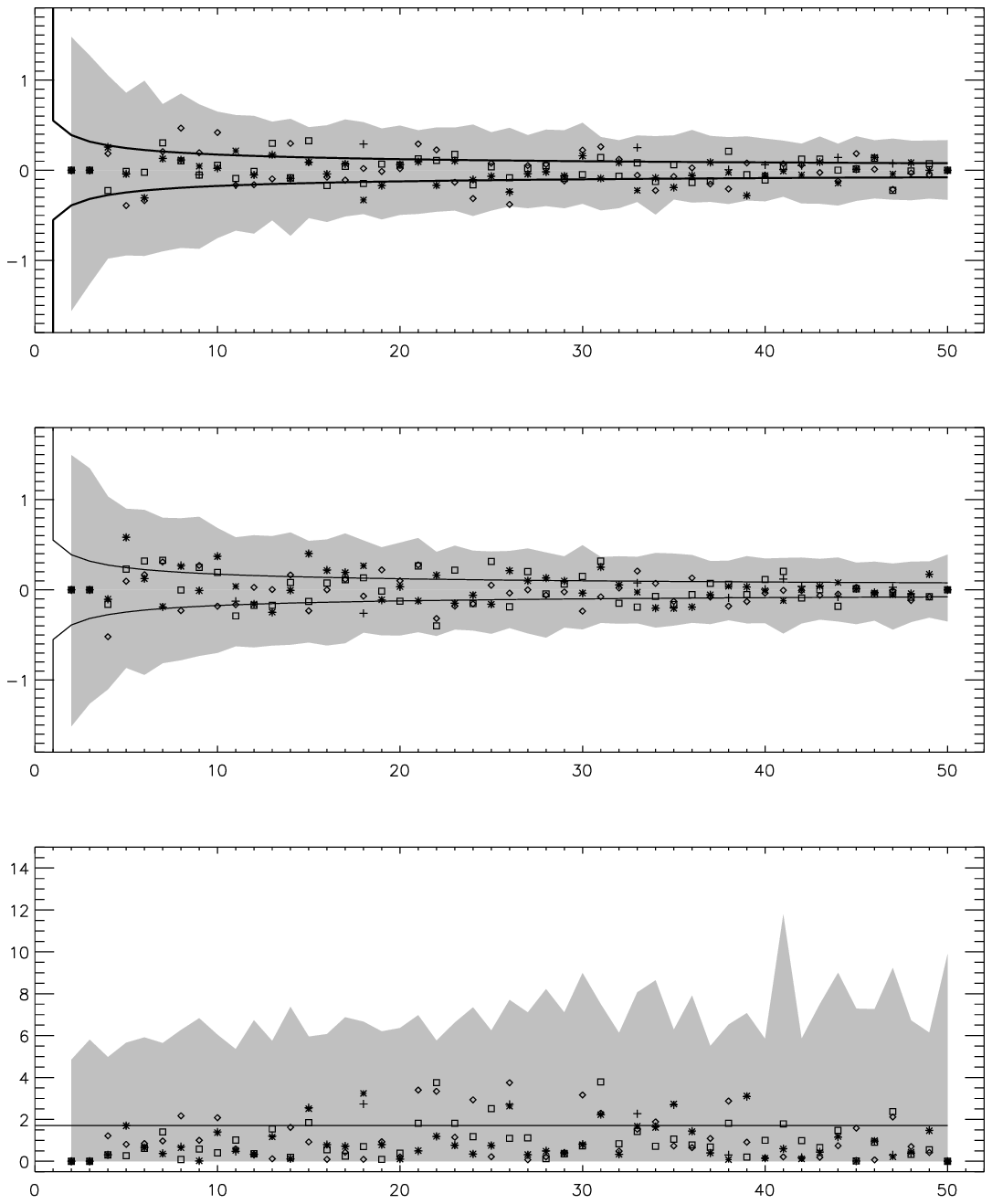,width=6.2cm,height=11cm}
\epsfig{file=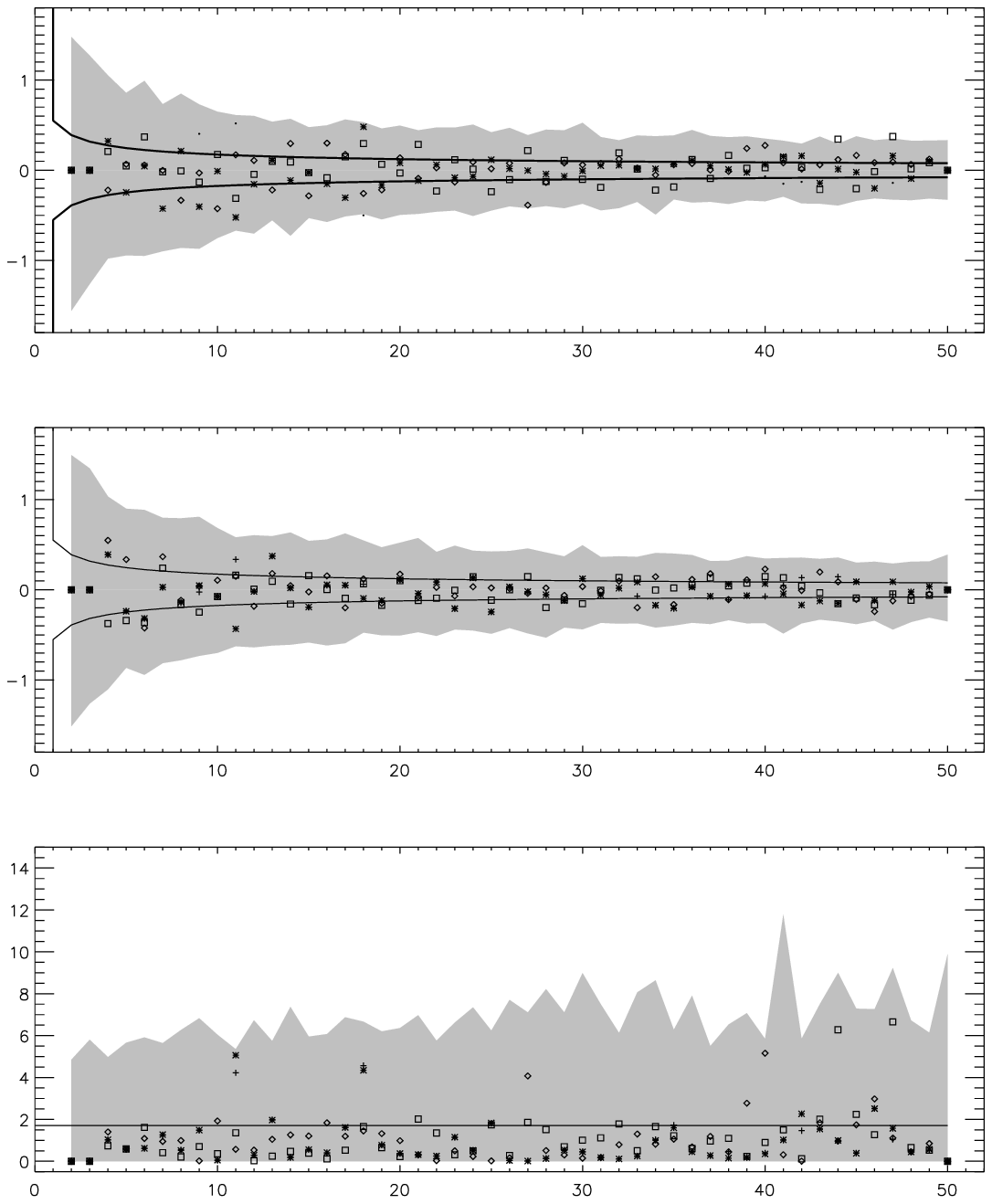,width=6.2cm,height=11cm}
\epsfig{file=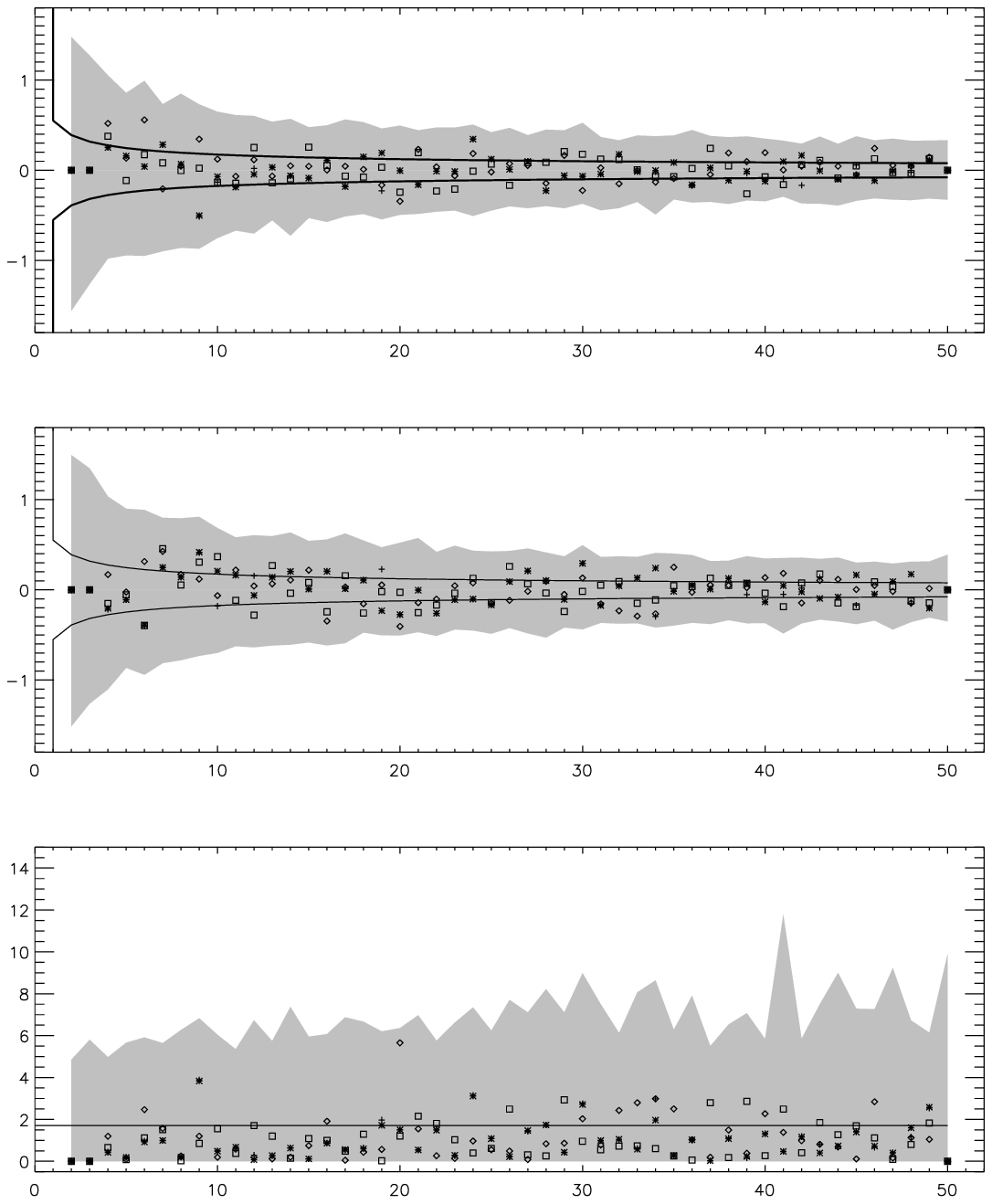,width=6.2cm,height=11cm}\\
\caption{The circular statistics for the
  cross correlation of phases between the ILC and the foreground maps
  (top 3 rows) and between the PCM and the foreground maps (lower 3
  rows). Each 3 rows (downwards) are $\Si$, $\Cs$ and $r^2$ statistics,
  respectively. The 3 columns from left to right are for $\Delta \l=0$, 1 and
  2, respectively. In each panel, the star sign marks ILC-Ka
  foreground cross correlation, the plus sign marks
  ILC-Q, the diamond sign marks ILC-V and the box sign ILC-W. The
  solid curves denote 1$-\sigma$ CL and the shaded areas cover 1000
  Gaussian realizations from Monte Carlo simulations. The right
  column are $\Si$, $\Cs$ and $r^2$ statistics between the PCM and
  foregrounds for $\Delta \l=0$, 1 and 2, respectively.       
}
\label{ilcpcm}
\end{figure*}

\begin{figure*}
\epsfig{file=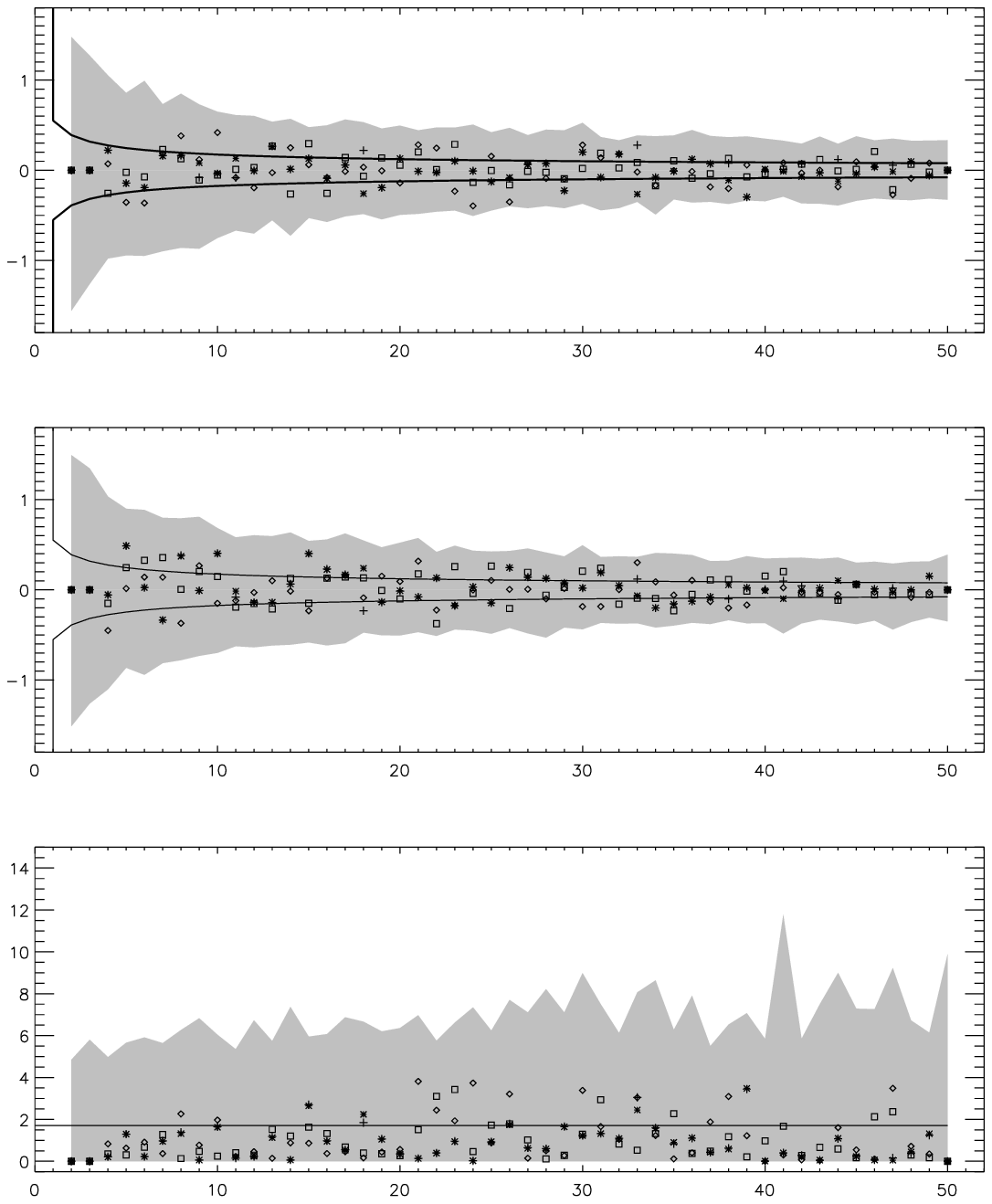,width=6.2cm,height=11cm}
\epsfig{file=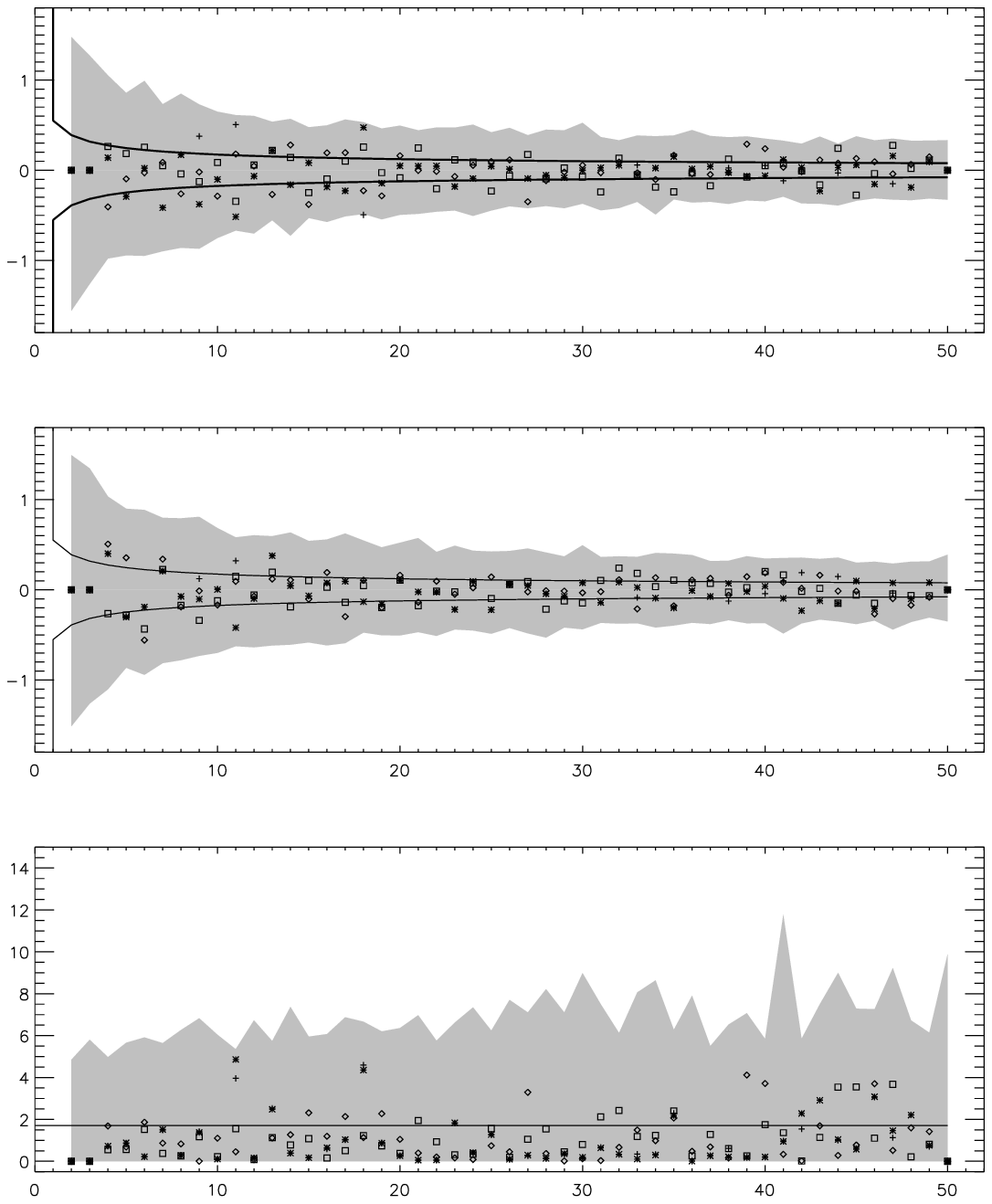,width=6.2cm,height=11cm}
\epsfig{file=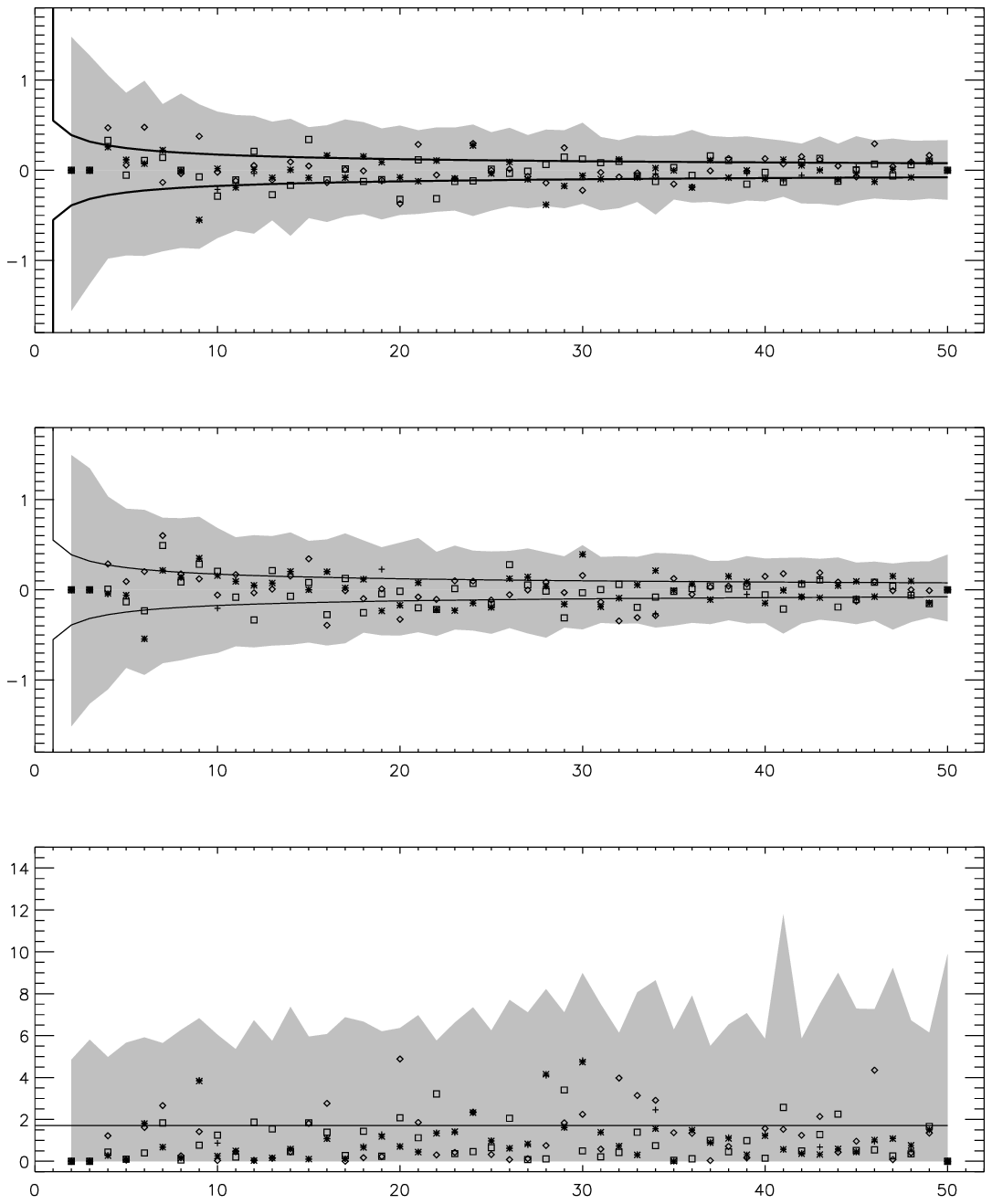,width=6.2cm,height=11cm}\\
\epsfig{file=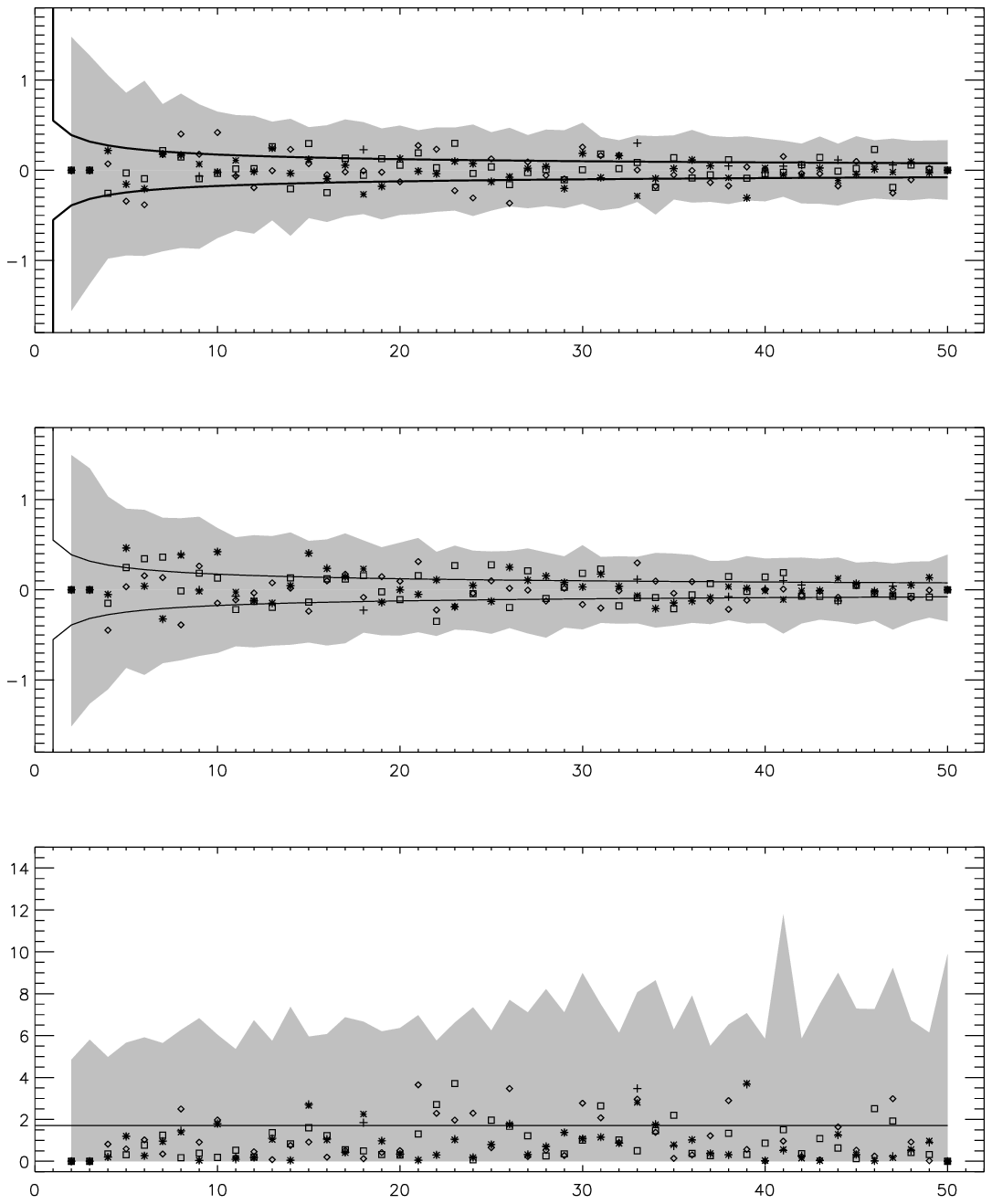,width=6.2cm,height=11cm}
\epsfig{file=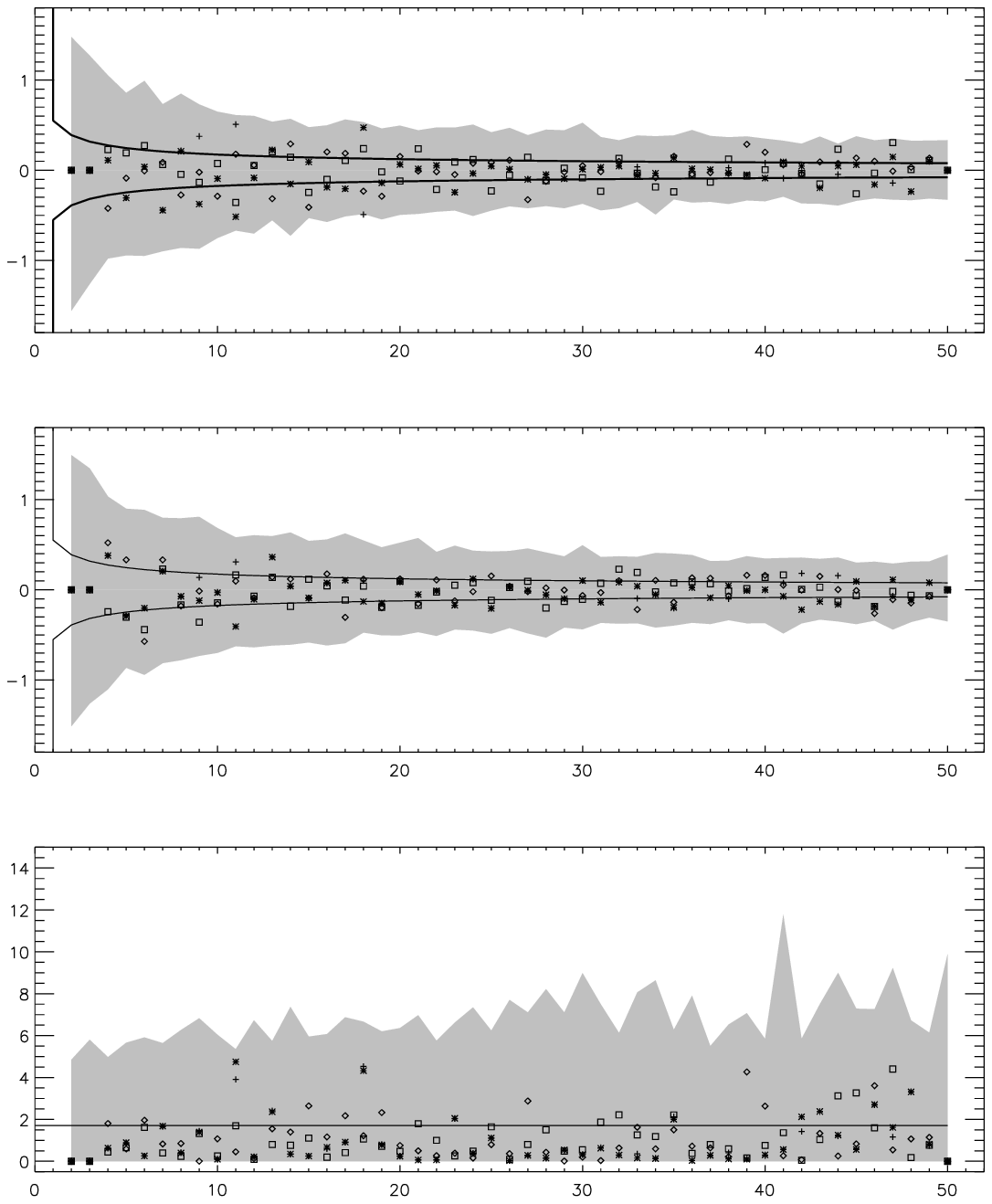,width=6.2cm,height=11cm}
\epsfig{file=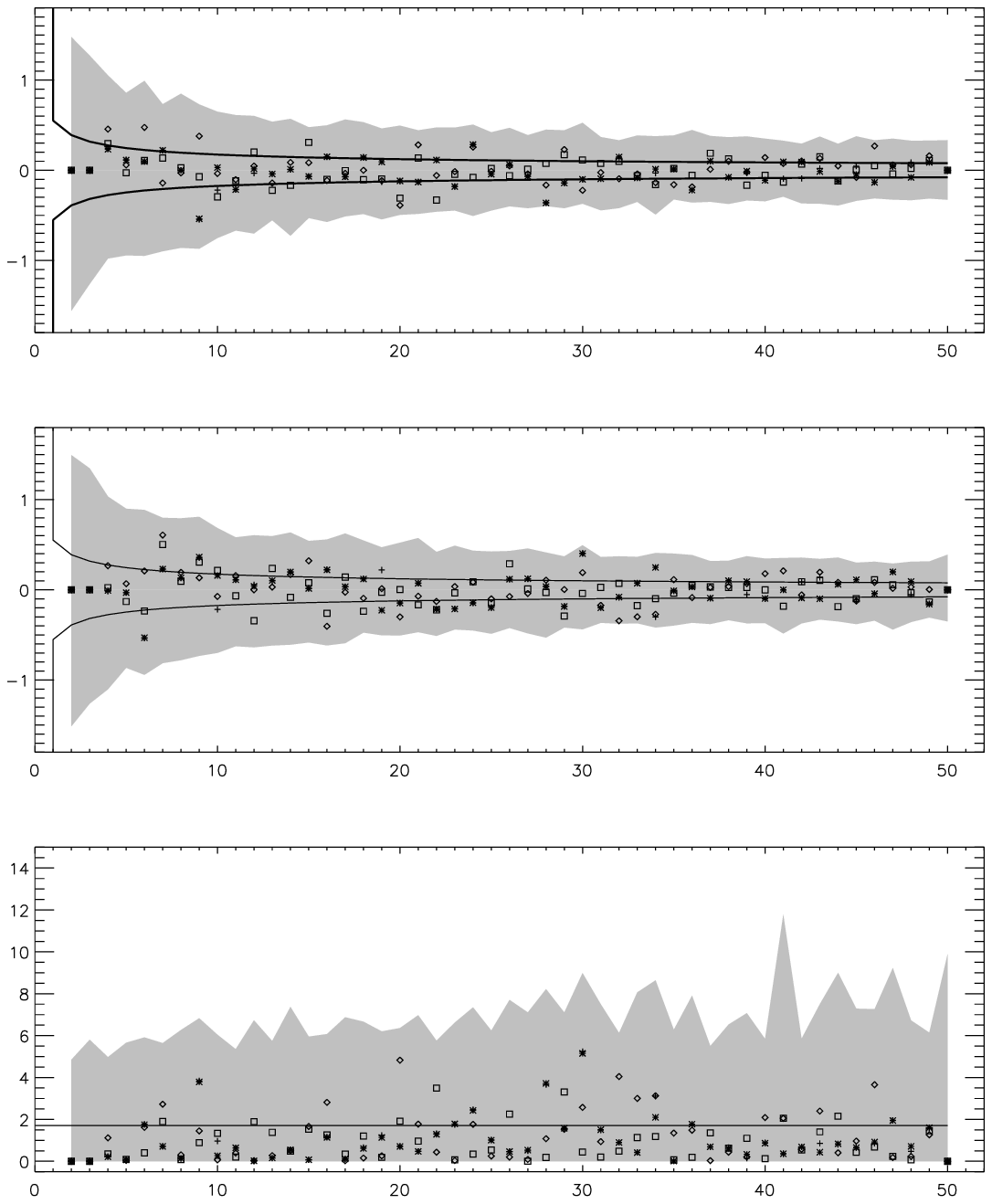,width=6.2cm,height=11cm}\\
\caption{The circular statistics for the
  cross correlation of phases between the TOH FCM and foreground maps
  (top 3 rows) and between the WFM and foreground maps (lower 3
  rows). The notation is the same as Fig.\ref{ilcpcm}.}
\label{fcmwfm}
\end{figure*}

In Fig.\ref{ilcpcm} and \ref{fcmwfm} we plot the $\Si$, $\Cs$ and $r^2$
statistics on the foreground-cleaned and the foreground maps. The top
3 rows of Fig.\ref{ilcpcm} is the statistics for ILC and foregrounds, in which
from left to right columns are for $\Delta \l =0$, 1 and 2,
respectively. One can see significant
cross correlations at multipoles $\l=6$, 8, 14-16, 21, 39 and 47 of
ILC with Q, V and W foregrounds at 95\% CL and at $\l=11$, 17, 26, 36
and 47 at 99\% CL. The bottom 3 rows of Fig.\ref{ilcpcm} is the statistics on
PCM and the foregrounds. PCM is produced without galactic cut-off and
any preliminary information about foreground properties
\citep{pcm}. The possible non-Gaussianity in the PCM is the residues
of bright point sources in the Galactic plane region covered by \wmap
Kp2 mask. These residues have pronounced cross correlations with the
synchrotron component as shown in Fig.\ref{Smap}. In Fig.\ref{fcmwfm}
on the top 3 rows we can also see clearly some multipoles with cross
correlations above 2$-\sigma$ and for $\l=30$ the correlation is above
99\% CL.  

To show the possible multipole range in which contamination of
non-Gaussian features can be important, in Fig.\ref{cosvar} we plot the
difference between the best-fit \wmap $\Lambda$CDM, ILC and TOH FCM
power spectra, normalized to the $\Lambda$CDM power. The intriguing
features appear at multipoles $\l=2$, 5, 7, $20-23$, 40, in which the
peaks of $\Delta \Cl/\Cl $ lies above 68\% CL. 

\begin{apjemufigure}
\hbox{\hspace*{-0.2cm}
\centerline{\includegraphics[width=1.\linewidth]{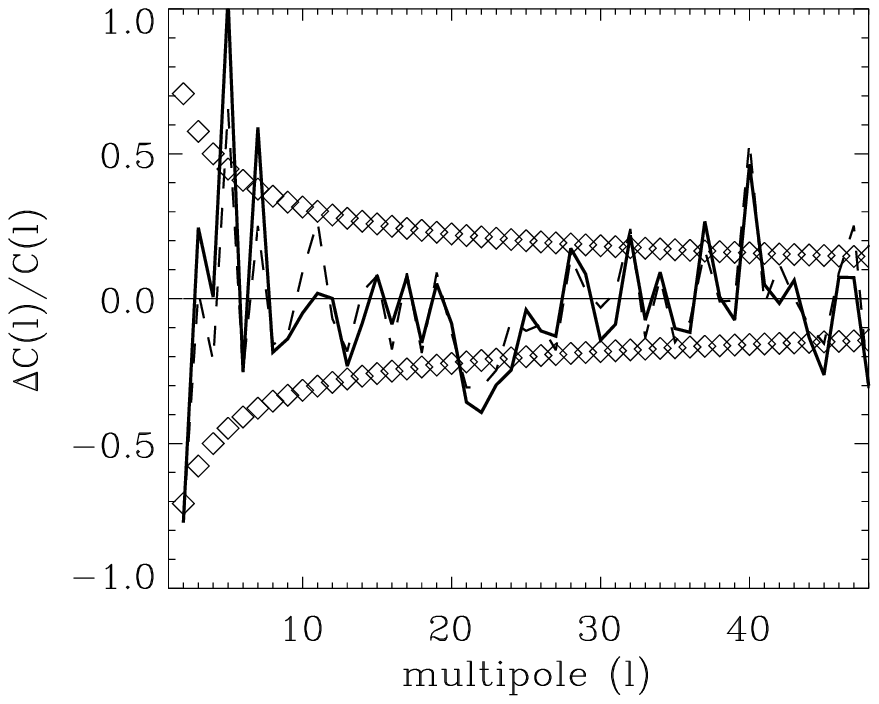}}}
\caption{The difference between the best-fit \wmap $\Lambda$CDM, ILC
  and TOH FCM power spectra, normalized to the $\Lambda$CDM power. The
  diamond shape curves represent the cosmic variance.}
  \label{cosvar} 
\end{apjemufigure}

\section{Conclusions and Discussions}
In this paper we raise the issue of how different the foreground maps
at different frequency bands. From each \wmap foreground component:
synchrotron, free-free and dust maps, we look for spectral-index
varying signals. By assuming that the spectral-index {\it invariant} part
has 100\% phase correlation (hence the same morphology) among
the different frequency bands, we develop a new method for extracting the
spectral-index varying signals, which correspond to perturbations in
phases. We have shown that this varying component has a characteristic
in phases: turning $\pi/2$ or $3\pi/2$ from the highly correlated
foregrounds.   

We further use the standard minimization of the variance to extract
the frequency independent signals from all the frequency bands. This
component can play a crucial role for the current blind method for
component separation in CMB studies. Our method is not only useful for
the foreground analysis on the \wmap data, it shall be very crucial
for the upcoming \planck data analysis.

\acknowledgements
We thank M. Tegmark et al. for providing
their processed maps. We thank A. Doroshkevich, M. Demianski and
P. R. Christensen for useful discussions. We acknowledge the use of
the Legacy Archive for Microwave Background Data Analysis
(LAMBDA). Support for LAMBDA is provided by the NASA Office of Space
Science. We also acknowledge the use of \healpix
\footnote{\tt http://www.eso.org/science/healpix/}
package \citep{healpix} to produce $\alm$ from the \wmap data and the
use of the \glesp package \citep{glesp} for data analyses and the
whole-sky figures.

\end{document}